\documentclass[]{aa}
\usepackage{graphicx}
\usepackage{adjustbox}
\usepackage{txfonts}
\usepackage{booktabs} 
\usepackage{caption}
\usepackage{xcolor}
\usepackage{placeins}

\makeatletter
\renewcommand*\aa@pageof{, page \thepage{} of \pageref*{LastPage}}
\makeatother
\usepackage{hyperref}
\hypersetup{
    colorlinks=true,
    linkcolor=blue,
    filecolor=magenta,      
    urlcolor=blue,
    citecolor=blue,
    }

\begin{document} 

\title{Multiple populations along the asymptotic giant branch: a Gaia+APOGEE study of 22 Galactic globular clusters}
\subtitle{}

\author{E. Dondoglio\inst{1,3}, A. P. Milone\inst{2,3}, A. F. Marino\inst{3}, G. Cordoni\inst{4}, M. V. Legnardi\inst{2}, T. Ziliotto\inst{5, 6}, R. Asa’d\inst{1}, A. Mastrobuono-Battisti\inst{2, 7}, F. Muratore\inst{2},  E. Bortolan\inst{2}, E. P. Lagioia\inst{7}, M. Tailo\inst{9}}

\institute{
Physics Department, American University of Sharjah, P.O. Box 26666, Sharjah, UAE \\ \email{edondoglio@aus.edu}
\and
Dipartimento di Fisica e Astronomia ``Galileo Galilei'', Univ. di Padova, Vicolo dell'Osservatorio 3, Padova, IT-35122
\and
Istituto Nazionale di Astrofisica - Osservatorio Astronomico di Padova, Vicolo dell’Osservatorio 5, Padova, IT-35122
\and
Research School of Astronomy and Astrophysics, Australian National University, Canberra, ACT 2611, Australia
\and
NSF’s NOIRLab, 950 North Cherry Avenue, Tucson, AZ 85719
\and
Space Telescope Science Institute, 3700 San Martin Dr, Baltimore, MD 21218, USA
\and 
Dipartimento di Tecnica e Gestione dei Sistemi Industriali, Università degli Studi di Padova, Stradella S. Nicola 3, I-36100, Italy
\and
South-Western Institute for Astronomy Research, Yunnan University, Kunming 650500, People's Republic of China
\and
Dipartimento di Fisica e Astronomia “Augusto Righi”, Universitá di Bologna, Via Gobetti 93/2, 40129 Bologna, Italy
}

\titlerunning{Multiple populations in the AGB with APOGEE}
\authorrunning{Dondoglio et al.}

\date{Received XXX / Accepted XXX}

\abstract{
We present a comprehensive investigation of multiple stellar populations along the asymptotic giant branch (AGB) in 22 Galactic globular clusters (GCs), exploiting APOGEE spectroscopy combined with a robust AGB selection based on Gaia color-magnitude diagrams. Using light-element abundances, including [C/Fe], [N/Fe], [Mg/Fe], and [Al/Fe], we successfully disentangle first- (1P) and second- (2P) populations along the AGB. We derive population fractions in the AGB for the largest sample of GCs to date, finding that the 1P fraction decreases with increasing cluster mass, in agreement with other evolutionary phases.
By directly comparing AGB and red giant branch (RGB) populations, we define a quantitative criterion to identify clusters affected by the AGB-manqué phenomenon. We find that in nine GCs the most chemically extreme 2P stars are significantly underrepresented along the AGB, indicating that a fraction of these stars fails to ascend this phase. In the remaining clusters, AGB and RGB abundance distributions are consistent within uncertainties. Our classification is in overall agreement with previous studies and provides the first spectroscopic characterization of AGB multiple populations in eight GCs.
Combining our measurements with literature data, we derive, for the first time, the radial distribution of AGB 2P stars in four clusters. While NGC\,5024 and $\omega$Centauri show trends consistent with the RGB, NGC\,2808 and NGC\,7078 may exhibit an unexpected increase of the AGB 2P fraction at large radii, opposite to the RGB stars.
We also present the first detailed spectroscopic characterization of anomalous
AGB populations in NGC\,6656 and $\omega$Centauri, i.e. of the populations
enhanced in heavy elements compared to the bulk of 1P and 2P.
In both clusters, anomalous stars show a much more pronounced signature of the AGB-manqué presence than 2P stars, with the fraction of the most Mg-poor and Al-rich AGB dramatically dropping compared to the RGB, possibly due to enhanced helium content and/or increased RGB mass loss.
Finally, we report the first detection of intrinsic iron inhomogeneities among 1P AGB stars in NGC\,5272, with a spread consistent with that observed along the RGB. This extends the presence of iron variations to the most evolved stellar phase studied so far.
}


\maketitle
\section{Introduction}
\label{sec:intro}

The mystery of chemical inhomogeneities in globular clusters (GCs) has persisted for several decades and is still far from being fully understood. The overall picture that has emerged from decades of intensive observational and theoretical work reveals that GCs host distinct stellar populations characterized by different light-element chemical patterns. Indeed, while a fraction of GC stars exhibit chemical compositions similar to those observed in halo field stars of their host galaxy (the so-called first-population stars; 1P), the remaining stars display light-element abundance patterns that are not observed in any other stellar environment in the Universe. These stars typically show enhanced He, N, Na, and Al, and depleted C, O, and Mg with respect to 1P stars, and are generally referred to as second-population (2P) stars. Moreover, 2P stars themselves are often composed by separate groups of stars, each characterized by chemical patterns that are more or less distant from the 1P chemistry \citep[see][for reviews]{bastian2018, milone2022}.

Over the years, additional features of multiple populations have been unveiled, adding further layers of complexity to the formation and evolution of these structures. In some GCs, such as NGC\,0104 (47Tucanae) and NGC\,2808, 2P are more centrally concentrated than 1P stars. However, this behavior is not universal: in several clusters (e.g., NGC6121 and NGC\,6752) no significant radial segregation is observed \citep[e.g., ][]{dondoglio2021, leitinger2023, jang2025}. Moreover, about one-fifth of Galactic GCs host, in addition to 1P and 2P stars, an extra group of stars enriched in iron, \textit{s}-process elements, and/or C+N+O. These stars are commonly referred to as anomalous, and clusters hosting them are classified as Type~II, in contrast to Type~I clusters, which only contain 1P and 2P stars \citep[][]{milone2017}. Interestingly, even the anomalous stars display light-element inhomogeneities, somewhat similar to those observed between 1P and 2P stars \citep[e. g.,][]{marino2011, yong2015, dondoglio2023}. Finally, 1P stars, traditionally considered to be chemically homogeneous, show a spread in photometric diagrams that is significantly larger than expected from observational errors alone \citep[][]{legnardi2022, legnardi2024}. This feature has been interpreted as an intrinsic [Fe/H] spread, as supported by several spectroscopic studies \citep[e.g.,][]{marino2019, lardo2023, latour2025}.

The observational properties of multiple populations appear to be shared across different evolutionary phases, from the bottom of the main sequence (MS) up to the horizontal branch (HB) \citep[e.g.,][]{milone2012, milone2017, dondoglio2021, dondoglio2022}. However, the situation becomes more complex when considering the asymptotic giant branch (AGB). Early spectroscopic studies of AGB stars in GCs primarily relied on molecular indices sensitive to CN and CH abundances \citep[e.g.,][]{mallia1978, norris1981, smith1989, smith1993}. These works established that, similarly to the red giant branch (RGB), AGB stars exhibit intrinsic chemical inhomogeneities. However, they also suggested that, in some GCs, the AGB population is underrepresented in the most N-rich and C-poor 2P stars. This behavior was later investigated using [Na/Fe] abundances, revealing that while some clusters display similar Na distributions along the RGB and AGB, others show a clear paucity of the most Na-rich 2P stars, with notable examples including NGC\,2808, NGC\,6266, and NGC\,6752. Subsequent spectroscopic studies exploiting additional abundance diagnostics, such as [N/Fe], [O/Fe], and [Al/Fe], broadly confirmed this dichotomy \citep[e.g.,][]{piliachowski1996, ivans2001, campbell2013, garcia2015, lapenna2015, lapenna2016, wang2016, wang2017, marino2017, gerber2018, mucciarelli2019}. Consistent results have also been obtained from photometric investigations, which revealed multiple populations along the AGB in several clusters and, again, an absence of the most chemically extreme 2P stars in some cases, while others show no significant differences between the RGB and AGB \citep[e.g.,][]{monelli2013, gruyters2017, lardo2017, lee2017, lagioia2021, jang2022}.

Why does this happen on the AGB? The commonly accepted interpretation is that some stars fail to ascend the AGB. This occurs when a star has a low envelope mass ($M_{\rm env}$) during its HB phase, which is insufficient to allow the star to expand and cool while evolving toward the AGB. This condition ($M_{\rm env} \lesssim 10^{\rm -2} M_{\rm \odot}$) is met by the hottest (hence the bluest) HB stars, which therefore skip the AGB phase. These objects are known as AGB-manqué \citep[][]{greggio1990, dorman1993}.
Several factors are expected to regulate the occurrence of AGB-manqué stars, including GC age, metallicity, initial He mass fraction, and RGB mass loss \citep[e.g.,][]{charbonnel2013, cassisi2014, charbonnel2016}. The latter two are of particular relevance in multiple populations, since 2P stars are observed to be He-enriched \citep[][]{milone2018} and to experience larger RGB mass-loss \citep[][]{tailo2020} than 1P stars. Indeed, within a given GC at fixed metallicity $Z$, a star is expected to reach higher effective temperatures ($T_{\rm eff}$), i.e. bluer locations along the HB (and be more likely to skip the AGB), if it has a higher He mass fraction and/or has experienced stronger mass loss during the RGB phase \citep[][]{dantona2002}.

Despite the growing body of evidence on multiple populations across different evolutionary phases, the AGB remains comparatively poorly explored. Most spectroscopic studies on this topic have focused on one or a small number of GCs, and a homogeneous analysis of the chemical abundance pattern of AGB stars over a large cluster sample is still lacking. As a consequence, the number of GCs for which robust spectroscopic constraints on AGB-manqué imprints are available remains limited. Moreover, the aforementioned observational features that have been extensively investigated in other evolutionary stages -- such as radial distributions, intrinsic [Fe/H] spreads among 1P stars, and the properties of anomalous populations -- have not yet been systematically explored along the AGB. In particular, while it is well established that anomalous stars host their own multiple population patterns, it is still unclear whether, and to what extent, they are affected by the AGB-manqué phenomenon.
\citet[][]{lagioia2021} through Hubble Space Telescope (HST) photometry, found that anomalous stars have a lower overall incidence along the AGB compared to the RGB, but a spectroscopic characterization of such subpopulations is still missing.

In this work, we combine Gaia photometry and Apache Point Observatory Galactic Evolution Experiment (APOGEE) spectroscopy to identify multiple populations among the AGB stars of 22 GCs: 47Tucanae, NGC\,0288, NGC\,1904, NGC\,2808, NGC\,3201, NGC\,4590, NGC\,5024, NGC\,5053, NGC\,5139 ($\omega$Centauri) NGC\,5272, NGC\,5904, NGC\,6121, NGC\,6171, NGC\,6205, NGC\,6218, NGC\,6254, NGC\,6341, NGC\,6397, NGC\,6656, NGC\,6809, NGC\,6838, and NGC\,7078. Section~\ref{sec:2} describes the dataset and the selection of the AGB and RGB stars. In Section~\ref{sec:3}, we identify 1P and 2P stars in both evolutionary phases, including the anomalous populations in $\omega$Centauri and NGC\,6656. Section~\ref{sec:4} presents the multiple population fractions and their radial distributions along the AGB, together with a comparison with the RGB. In Section~\ref{sec:5}, we establish a quantitative criterion to assess whether a given GC exhibits AGB-manqué signatures. Section~\ref{sec:6} reports the first detection of iron inhomogeneities among 1P AGB stars in NGC\,5272. Finally, Section~\ref{sec:7} summarizes the main results of this work.

\section{Dataset}
\label{sec:2}

\begin{figure}
\includegraphics[width=8.5cm, clip, trim={ 0cm 14.7cm 0cm 0cm}]{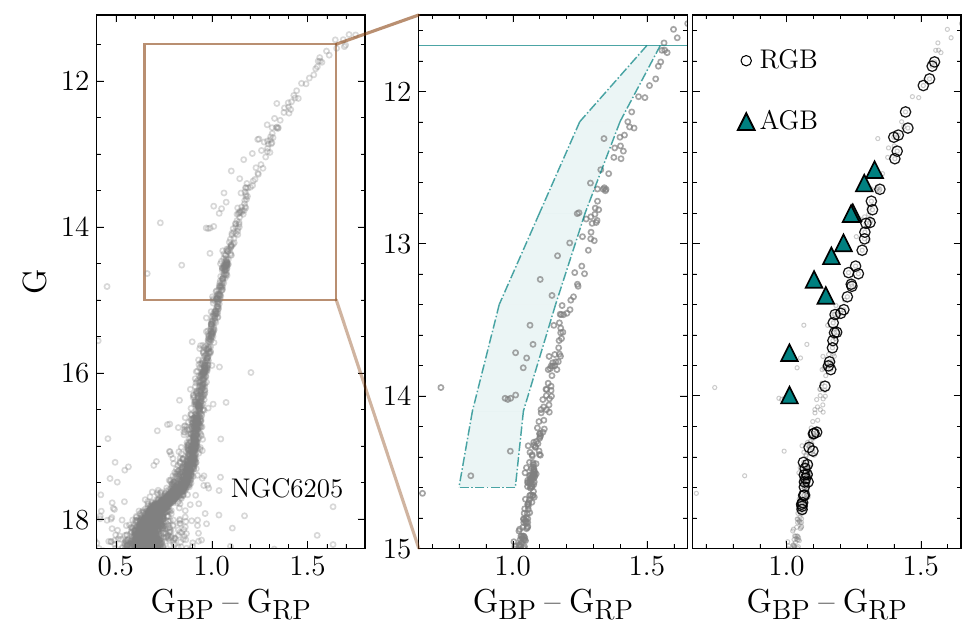}
\caption{{\it{Left:}} $G$ vs. $G_{\rm BP}$-$G_{\rm RP}$ CMD of NGC\,6205 members. The brown rectangle highlights the CMD area used for AGB stars selection. {\it{Middle:}} zoom-in of the CMD, in which the teal region indicates the area where we selected our AGB sample. The horizontal line indicates the magnitude threshold above which RGB and AGB sequence are not sharply separated. {\it{Right:}} same as middle panel but highlighting the RGB (open black dots) and AGB (teal triangles) stars cross-matched with APOGEE.}
\label{fig:sel_agb}
\end{figure}

To carry out our analysis, we combine public photometric and spectroscopic information for 22 Galactic GCs.
First, to identify a sample of cluster members in the AGB phase, we exploit Gaia photometry from Data Release 3 \citep[][]{gaia2023}, following a widely adopted approach in Gaia-based GC analyses \citep[e.g.,][]{jang2022}. Briefly, within a GC field we retain only the sources with high-quality photometry, as quantified by the renormalized unit weight error ({\tt{RUWE}}). Cluster members are then identified as stars with parallaxes and proper motions consistent with those of the inspected GC. The resulting catalog is corrected for differential reddening following the procedure outlined by \citet[][see also \citealt{legnardi2023}]{milone2012}. As an example, the left panel of Figure~\ref{fig:sel_agb} displays the $G$ vs. $G_{\rm BP}$--$G_{\rm RP}$ color-magnitude diagram (CMD) of NGC\,6205 members.

AGB stars are selected by eye based on their position in the CMD. Here, the brightest portion of the AGB sequence tends to merge with the RGB, making the two evolutionary phases indistinguishable. For this reason, we exclude from our analysis the bright magnitude range in which the two sequences are not sharply separated. The middle panel of Figure~\ref{fig:sel_agb} shows a zoom-in of the CMD around the region populated by AGB stars (brown box). The teal shaded region indicates the portion of the CMD populated by AGB stars.  We interrupt our selection at $G =$ 11.7 mag (teal horizontal line), as the two sequences are no longer well separated beyond this limit.

Once a reliable sample of AGB stars has been identified, we cross-match the Gaia photometry with spectroscopy from the APOGEE Data Release 17 \citep[DR17;][]{abdurrof2022}. This survey provides abundances of a large number of chemical species for approximately 700,000 stars, derived from H-band spectra. In this work, we specifically use the [C/Fe], [N/Fe], [O/Fe], [Mg/Fe], [Al/Fe], and [Fe/H] measurements. We restrict our analysis to bright sources with signal-to-noise ratios larger than 70 and exclude stars flagged with {\tt{ASCAPFLAG = STAR\_BAD}}\footnote{DR17 documentation: {\url{https://www.sdss4.org/dr17/}}.}
. A more detailed description of our selection criteria is provided in \citet[][]{dondoglio2025}.
In addition, we apply a further selection criterion based on stellar radial velocities measured by APOGEE. Specifically, we retain only stars whose radial velocities are clustered around the mean GC radial velocity, as reported by \citet[][]{baumgardt2018}.

The right panel of Figure~\ref{fig:sel_agb} illustrates with teal triangles the NGC\,6205 AGB stars with available chemical abundances. Black dots represent RGB stars within the APOGEE dataset.
The $G$ vs. $G_{\rm BP}$--$G_{\rm RP}$ CMDs and the selected RGB and AGB samples for all analyzed GCs are shown in Appendix~\ref{sec:ap1}.

\section{Multiple populations: AGB vs. RGB}
\label{sec:3}

In this Section, we identify the chemically different populations harbored in our sample of GCs among both RGB and AGB stars. First, we consider Type I clusters, for which we disentangle 1P and 2P stars in Section~\ref{sec:3.1}. Then, we investigate separately the Type II clusters NGC\,6656 and $\omega$Centauri in Section~\ref{sec:3.2} and~\ref{sec:3.3}, respectively, isolating their 1P, 2P and anomalous stars.

\subsection{First and second population in Type I globular clusters}
\label{sec:3.1}

\begin{figure}
\includegraphics[width=9cm, clip, trim={ 9cm 0cm 0cm 1cm}]{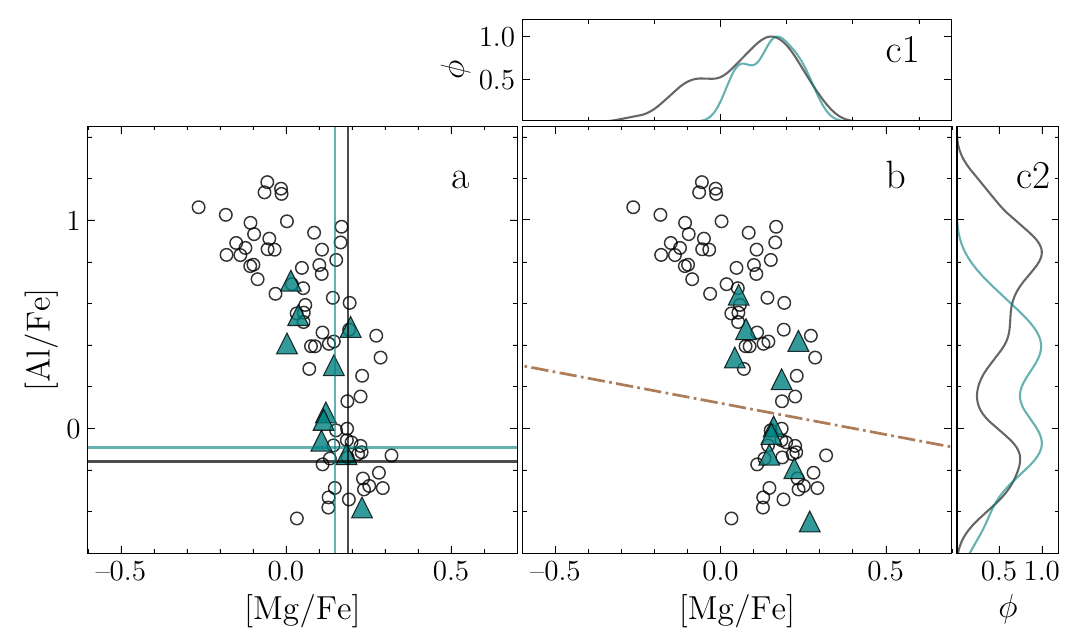}
\caption{{\it{Panel a:}} [Al/Fe] vs. [Mg/Fe] of RGB (black dots) and AGB (teal triangles) stars of NGC\,6205. The black and teal lines represent the median [Mg/Fe] and [Al/Fe] of the 1P stars in the RGB and AGB. {\it{Panel b:}} same diagram corrected for offsets (see text for details), with the brown dot-dashed line separating 1P and 2P stars (below and above, respectively). {\it{Panels c1 and c2:}} [Mg/Fe] and [Al/Fe] kernel density distribution, respectively, of the two groups of stars.}
\label{fig:mg_al_6205}
\end{figure}

To disentangle 1P and 2P stars, we exploit the [Mg/Fe] vs. [Al/Fe] diagram, as Figure~\ref{fig:mg_al_6205} illustrates for our template GC NGC\,6205. Both RGB and AGB stars (black dots and teal triangles, respectively) show 1P stars forming a separate distribution below [Al/Fe]$\sim$0.1 dex\footnote{We know they are 1P stars from \citet[][]{dondoglio2025}, who cross-matched 1P and 2P stars identified from photometric diagrams with the APOGEE dataset, finding that they cluster around this threshold, well separated from the bulk of Al-richer 2P stars (see their Figure~A4).}.
We notice that 1P in the RGB and AGB phase display a slight offset in their Mg and Al amount, as shown by their different [Mg/Fe] and [Al/Fe] median abundance (indicated by vertical and horizontal lines, respectively). To correct for this offset, we shift the AGB abundances to match the 1P median Mg and Al of RGB stars.
Among our sample of GCs, we measure [Mg/Fe] offset between $\sim$--0.08 and 0.07 dex, and between $\sim$--0.07 and 0.15 dex for [Al/Fe]. We dedicate Appendix~\ref{sec:ap3} to a more in-depth analysis of such offset, which are likely not reflecting intrinsic differences.
Panel b of Figure~\ref{fig:mg_al_6205} represents the [Al/Fe] vs. [Mg/Fe] corrected by these offsets, where the dot-dashed brown line separates the 1P (below) from the 2P (above). 2P stars of NGC\,6205 span a wide Mg and Al interval in the RGB, as well established in the literature \citep[e.g.][]{johnson2012}, ranging from stars with ratios closer to 1P values, to the most extreme 2P stars, characterized by with a [Al/Fe] enhancement up to $\sim$1.4 dex and [Mg/Fe] depletion of $\sim$0.5 dex compared to the 1P.
However, we notice a different behavior among the AGB stars. While this group of stars still shows the presence of 2P stars, Figure~\ref{fig:mg_al_6205} proves that no 2P star above [Al/Fe]$\sim$0.65 dex and below [Mg/Fe]$\sim$--0.05 dex is present in this evolutionary phase, as also supported by the [Mg/Fe] and [Al/Fe] kernel density distributions (panel c1 and c2), which display a clear lack of the most Mg-poor and Al-rich stars in the AGB. Our result demonstrates the imprint of the AGB-manqué phenomenon in NGC\,6205, as suggested by previous works \citep[][]{piliachowski1996, johnson2012}, and establish the APOGEE-based [Al/Fe] vs. [Mg/Fe] diagram as a reliable tool to investigate the AGB-manqués.

For five GCs on our sample, namely 47Tucanae, NGC\,6121, NGC\,6171, NGC\,6218, and NGC\,6838, the [Al/Fe] vs. [Mg/Fe] does not constitute an effective tool to isolate 1P and 2P stars. These clusters are among the most metal-rich in this study, and it is well known that at high metallicities star-to-star Mg and Al differences are very small if not negligible \citep[e.g.,][]{pancino2017, meszaros2020}. For that, we exploit the C and N abundances in the APOGEE dataset.
However, particular care must be taken when using the [C/Fe] and [N/Fe] abundance, as their surface abundances are affected by mixing episodes occurring after the first dredge-up, which produce a decrease in carbon and an increase in nitrogen with stellar luminosity above the RGB bump. This effect can hinder a proper multiple populations identification over a wide magnitude range.

\begin{figure}
\includegraphics[width=9cm, clip, trim={ 6.5cm 0cm 0cm 0cm}]{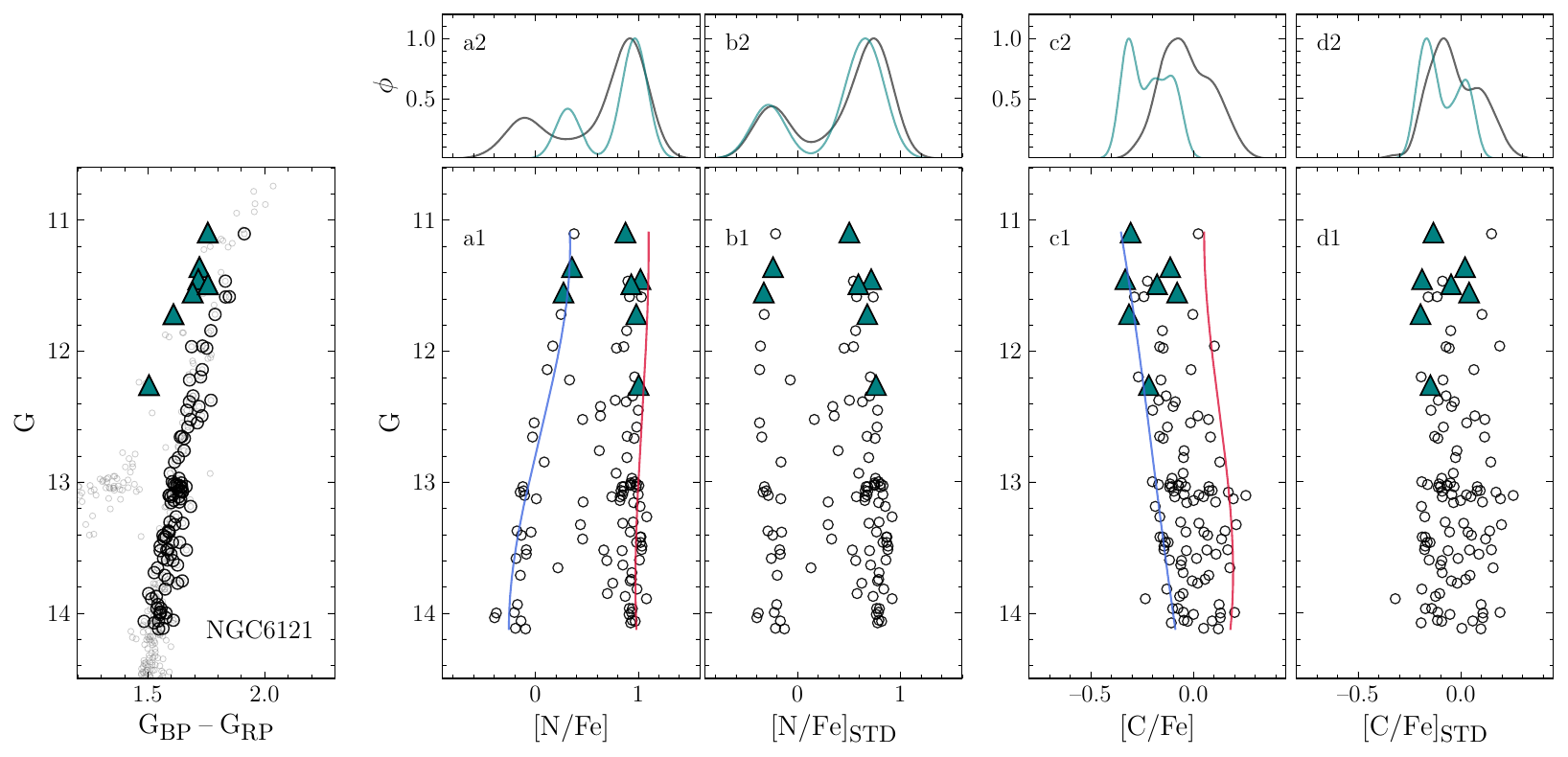}
\caption{
{\it{Panels a1 and a2:}} $G$ vs. [N/Fe] diagram and [N/Fe] kernel density distribution of RGB and AGB stars in NGC\,6121.
{\it{Panels b1 and b2:}} same but referred to the verticalized [N/Fe]$_{\rm STD}$ (see text for details). 
{\it{Panels c1, c2, d1, and d2:}} same as before but applied to the [C/Fe] abundances.}
\label{fig:delta_n}
\end{figure}

To correct for this effect, we follow the procedure outlined in Figure~\ref{fig:delta_n} by using NGC\,6121 as a template, where we show the $G$ vs. [N/Fe] abundance (panel a1) of RGB and AGB stars.
Two well-separated sequences are visible, especially at magnitudes fainter than $G \sim 13$ mag (which roughly corresponds to the RGB bump), where 1P and 2P stars cluster around [N/Fe] $\sim$--0.2 and $\sim$1.0 dex, respectively. Toward brighter magnitudes, the two sequences can still be traced, but their separation decreases: the 1P sequence shows an increase in [N/Fe] of $\sim$0.4 dex along the whole magnitude range. Intriguingly, the 2P stars do not exhibit a similar behavior, remaining approximately constant.

To correct the [N/Fe] abundances for this luminosity-dependent effect, we adopt a “verticalization” approach similar to that introduced by \citet{milone2015}  to better distinguish  multiple populations in HST photometric diagrams. We compute the 4th and 96th percentiles of the [N/Fe] distribution in different bins of magnitude, which are then fitted with splines to define the nitrogen-poor and nitrogen-rich boundaries ($f_{\rm poor}$ and $f_{\rm rich}$) of the distribution, indicated with the blue and red lines, respectively. Following the formalism described in \citet[][see their Section~3.2]{milone2017},we derive the verticalized $\Delta_{\rm [N/Fe]}$ distribution as follows:

\begin{equation}
    \Delta_{\rm [N/Fe]} = W_{\rm [N/Fe]} {{\rm [N/Fe]} - f_{\rm poor} \over{(f_{\rm rich} - f_{\rm poor})}},
\end{equation}
where $W_{[\rm N/Fe]}$ is the average width of the [N/Fe] distribution below the RGB bump, where the luminosity-dependent trend is negligible. This parameter aims at normalizing $\Delta_{\rm [N/Fe]}$ such that its width is comparable to the [N/Fe] one below $G = 13$ mag. We then define the standardized [N/Fe], [N/Fe]$_{\rm STD}$, as

\begin{equation}
    {\rm [N/Fe]}_{\rm STD} = {\rm [N/Fe]}_{\rm 0} + \Delta_{\rm [N/Fe]},
\end{equation}
where $[N/Fe]_{\rm 0}$ is the average 1P [N/Fe] below the RGB bump. By adding this quantity, we ensure that [N/Fe]$_{\rm STD}$ average 1P and 2P values matches the one of the [N/Fe] distribution below the bump.
Panel b1 displays the corrected G vs. [N/Fe]$_{\rm STD}$ plot, where 1P and 2P sequence run parallel around the same average [N/Fe]$_{\rm STD}$ than [N/Fe] below the RGB bump. By comparing the [N/Fe] and [N/Fe]$_{\rm STD}$ kernel density distribution (panel a2 and b2, respectively), the latter exhibit a much sharper and more clear-cut two-peak separation than the former. We then repeated this procedure for the G vs. [C/Fe] (panel c1) trend. Panel d1 illustrates the resulting G vs. [C/Fe]$_{\rm STD}$ diagram, while the kernel distributions of [C/Fe] and [C/Fe]$_{\rm STD}$ are displayed in panel c2 and d2, respectively. We then derive [C/Fe]$_{\rm STD}$ and [N/Fe]$_{\rm STD}$ also for 47Tucanae, NGC\,6171, NGC\,6218, and NGC\,6838.

\begin{figure*}
\includegraphics[width=17.5cm, clip]{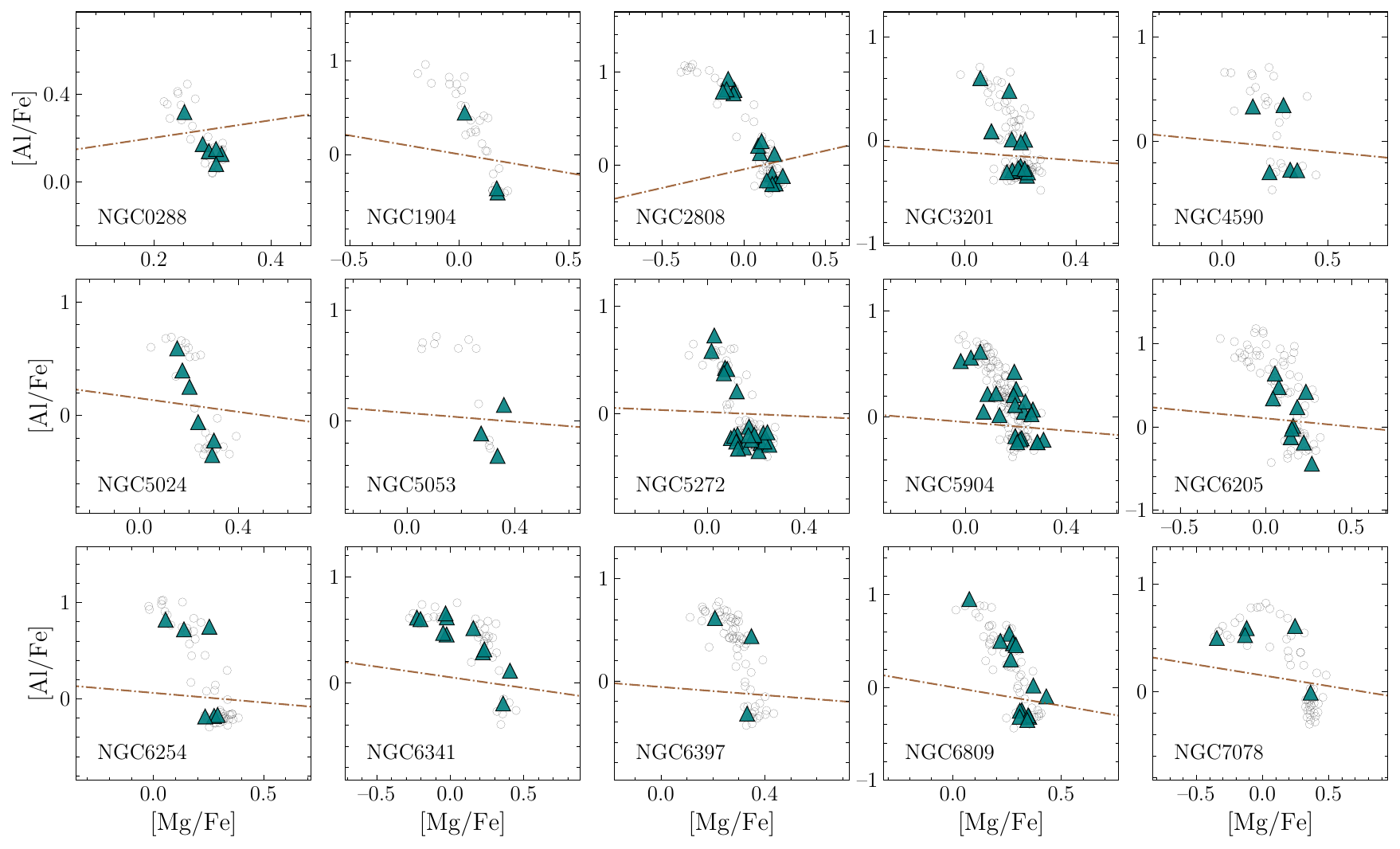}
\includegraphics[width=17.5cm, clip, trim={-0.35cm 0cm 0cm 0cm}]{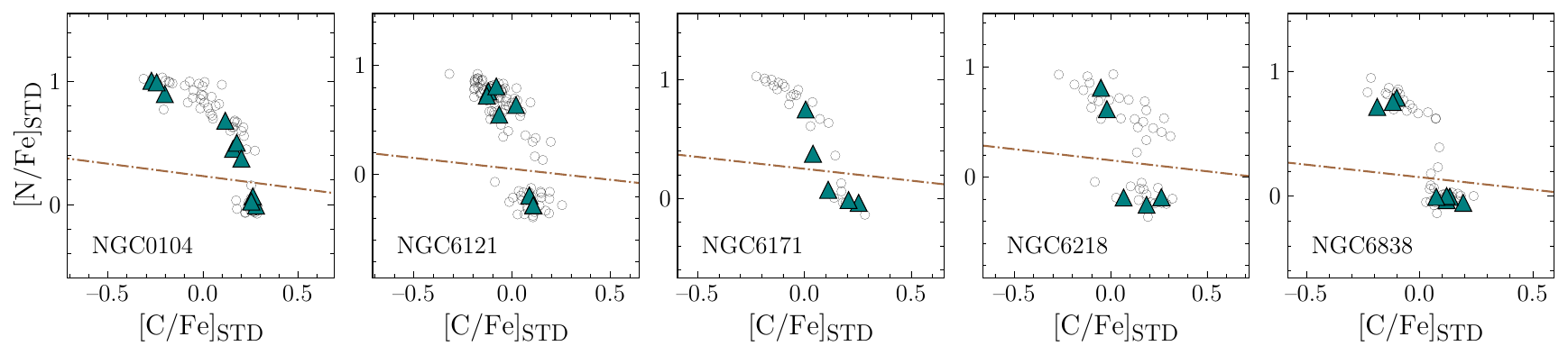}
\caption{[Mg/Fe] vs. [Al/Fe] diagram of 15 GCs, corrected for offsets as in Section~\ref{sec:2} (first three rows), and [N/Fe]$_{\rm STD}$ vs. [C/Fe]$_{\rm STD}$ of 47Tucanae, NGC\,6121, NGC\,6171, NGC\,6218, and NGC\,6838. The notation is the same as Figure~\ref{fig:mg_al_6205}.}
\label{fig:mg_al}
\end{figure*}

In Figure~\ref{fig:mg_al}, we show in the first three rows the [Al/Fe] vs. [Mg/Fe] diagram of 15 GCs, where the abundances of AGB stars have been corrected for offsets as done in Figure~\ref{fig:mg_al_6205}. The bottom row illustrates the [N/Fe]$_{\rm STD}$ vs. [C/Fe]$_{\rm STD}$ for the other five Type I clusters.
In all these GCs, a distinct, chemically uniform, clump of stars centered at the lowest y-axis values is visible, with such stars also displaying among the highest x-axis amount, corresponding to 1P stars. The brown lines separate, in each plot, the 1P from the 2P. By visual inspection, we can observe the AGB-manqué phenomenon (i.e., lack of the most chemically-extreme 2P stars) in NGC\,0288, NGC\,1904, NGC\,2808, NGC\,4590, and NGC\,5053 in the [Al/Fe] vs. [Mg/Fe] diagram, while for the [N/Fe]$_{\rm STD}$ vs. [C/Fe]$_{\rm STD}$ diagram NGC\,6171 represents the most evident case, but also NGC\,6121 displays a small clump of the most extreme 2P stars ([C/Fe]$_{\rm STD}$ $<$--0.2) that has no AGB representation. In Section~\ref{sec:5}, we will investigate in details the AGB-manqué features in these GCs.

\subsection{Populations in the Type II NGC\,6656}
\label{sec:3.2}

NGC\,6656 is one of the most extensively studied Type~II GC, with its anomalous population exhibiting enhancements in iron, $s$-process elements, and total C$+$N$+$O abundance \citep[e.g.][]{marino2009, marino2011, mckenzie2022, dondoglio2025}.

\begin{figure*}
\includegraphics[width=17.0cm, clip, trim={ 0cm 0cm 0cm 13.5cm}]{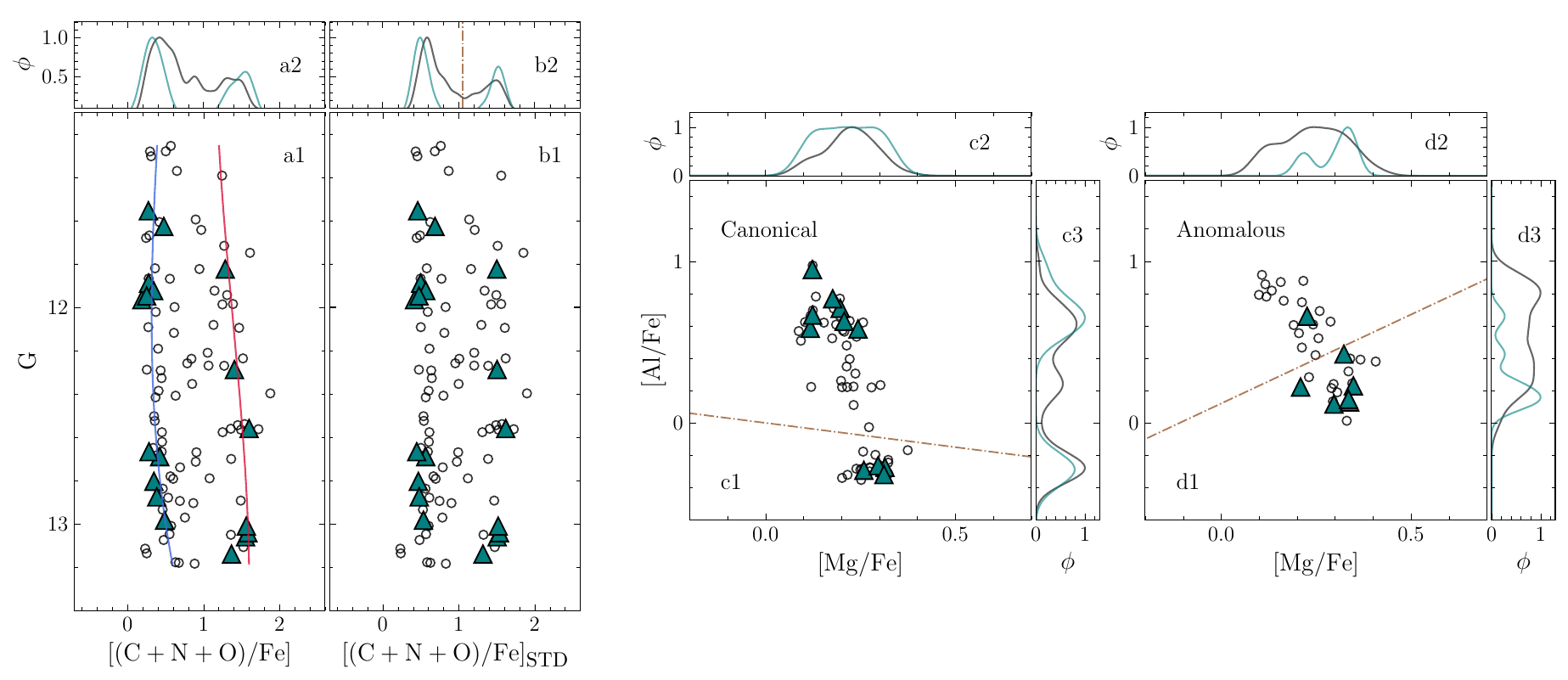}
\caption{
{\it{Panel a1:}} G vs. [(C$+$N$+$O)/Fe] diagram of RGB and AGB stars in NGC\,6656. Blue and red lines indicate the CNO-poor and CNO-rich boundaries (see text for details).
{\it{Panel a2:}} [(C$+$N$+$O)/Fe] kernel density distribution of RGB (black) and AGB (teal) stars.
{\it{Panels b1 e b2:}} same as a1 and a2 but for the verticalized [(C$+$N$+$O)/Fe]$_{\rm STD}$. The vertical brown dot-dashed line in panel b2 separate canonical (left) and anomalous (right) stars.
{\it{Panels c1 and d1:}} [Al/Fe] vs. [Mg/Fe] of canonical and anomalous stars, respectively. The brown lines separate 1P and 2P stars and AI and AII stars.
{\it{Panels c2 and c3:}} kernel density distribution of [Mg/Fe] and [Al/Fe] for canonical RGB and AGB stars.
{\it{Panels d2 and d3:}} same as c2 and c3 but for anomalous RGB and AGB stars.
}
\label{fig:ngc6656}
\end{figure*}

To isolate 1P and 2P stars in a manner consistent with our analysis of Type~I GCs, we must first distinguish between canonical and anomalous stars. To this end, we exploit the difference in the total C$+$N$+$O between the bulk of canonical and anomalous. Panel a1 of Figure~\ref{fig:ngc6656} shows the $G$ vs. [(C$+$N$+$O)/Fe] distribution for RGB and AGB stars. The [(C$+$N$+$O)/Fe] ratio does not exhibit a strong dependence on magnitude, as the increase in N caused by mixing processes is largely compensated by a decrease in C \citep[e.g.,][]{charbonnel1998}. Nevertheless, the CNO-poor and CNO-rich boundaries (blue and red lines, respectively), derived as in Figure~\ref{fig:delta_n}, display a slight decreasing trend with $G$. To correct for this effect, we apply equation 1 to the C$+$N$+$O abundance, obtaining the verticalized [(C$+$N$+$O)/Fe]$_{\rm STD}$ shown in panel~b1. Panels a2 and b2 compare the RGB and AGB kernel distribution of [(C$+$N$+$O)/Fe] and [(C$+$N$+$O)/Fe]$_{\rm STD}$, respectively, showing how the latter provides a sharper separation between the two peaks located at 0.5 and 1.6 dex, which represent the canonical and the CNO-enhanced anomalous population, respectively. These two groups of stars are separated by the vertical brown line in panel b2, corresponding to the minimum of the [(C$+$N$+$O)/Fe]$_{\rm STD}$ distribution between the two peaks: we consider as canonical (anomalous) stars the ones at the left (right) of this line.

Panel~c1 of Figure~\ref{fig:ngc6656} shows the [Al/Fe] vs. [Mg/Fe] diagram for the canonical populations, after correcting for offsets as in Figure~\ref{fig:mg_al_6205}. Similarly to Figure~\ref{fig:mg_al}, 1P stars produce a separate, Mg-rich and Al-poor clump (see the [Mg/Fe] and [Al/Fe] kernel density distribution in panels~c2 and~c3, respectively), while the stars above the brown line belong to the 2P.
In panel~d1, we illustrate the same plot but for the anomalous stars, where light-element inhomogeneities are also present, as proven by the observed Mg-Al anticorrelation similar to that observed in 1P and 2P stars, albeit with a smaller y-axis extent and spanning a [Al/Fe] skewed toward higher values \citep[as firstly detected by][]{marino2011}. This latter behavior is shared by anomalous stars in other Type II GCs \citep[like NGC\,1851 and NGC\,5286][]{carretta2011, marino2015}, likely indicating peculiar pollution mechanisms that influenced the composition of the medium from which they formed.
Indeed, the corresponding kernel density distributions (panels~d2 and~d3) reveal elongated Mg and Al distribution, supporting the presence of multiple populations among anomalous stars. In line with previous studies of anomalous subpopulations \citep[][]{dondoglio2023, dondoglio2026}, we do not label them as 1P and 2P (as they likely experienced different formation mechanisms), but we designate the Mg-rich and Al-poor anomalous population as AI, and the more extended sequence toward lower Mg and higher Al as AII. The brown dot-dashed line separates these two populations.

The AGB canonical stars cover a range on the [Al/Fe] vs. [Mg/Fe] similar to the RGB ones, thus suggesting the lack of AGB-manqué. On the other hand, the anomalous AGB stars exhibit remarkable differences to the RGB: all the anomalous AGB stars except one belong to the AI population, showing a severe lack of AII stars above [Al/Fe]$\sim$0.7 dex and below [Mg/Fe]$\sim$0.2 (see the also the distributions in panels~d2 and~d3).

\subsection{Populations in the Type II $\omega$Centauri}
\label{sec:3.3}

$\omega$~Centauri is the most chemically complex Galactic GC, hosting a prominent and variegated anomalous population characterized by a large spread in [Fe/H] (exceeding 1 dex), and strong enhancements in $s$-process elements and in the total C$+$N$+$O \citep[e.g.,][]{norris1995, johnson2010, marino2011a, nitschai2023, wang2026}.

\begin{figure*}
\includegraphics[width=17.0cm, clip, trim={ 0cm 0cm 0cm 13.8cm}]{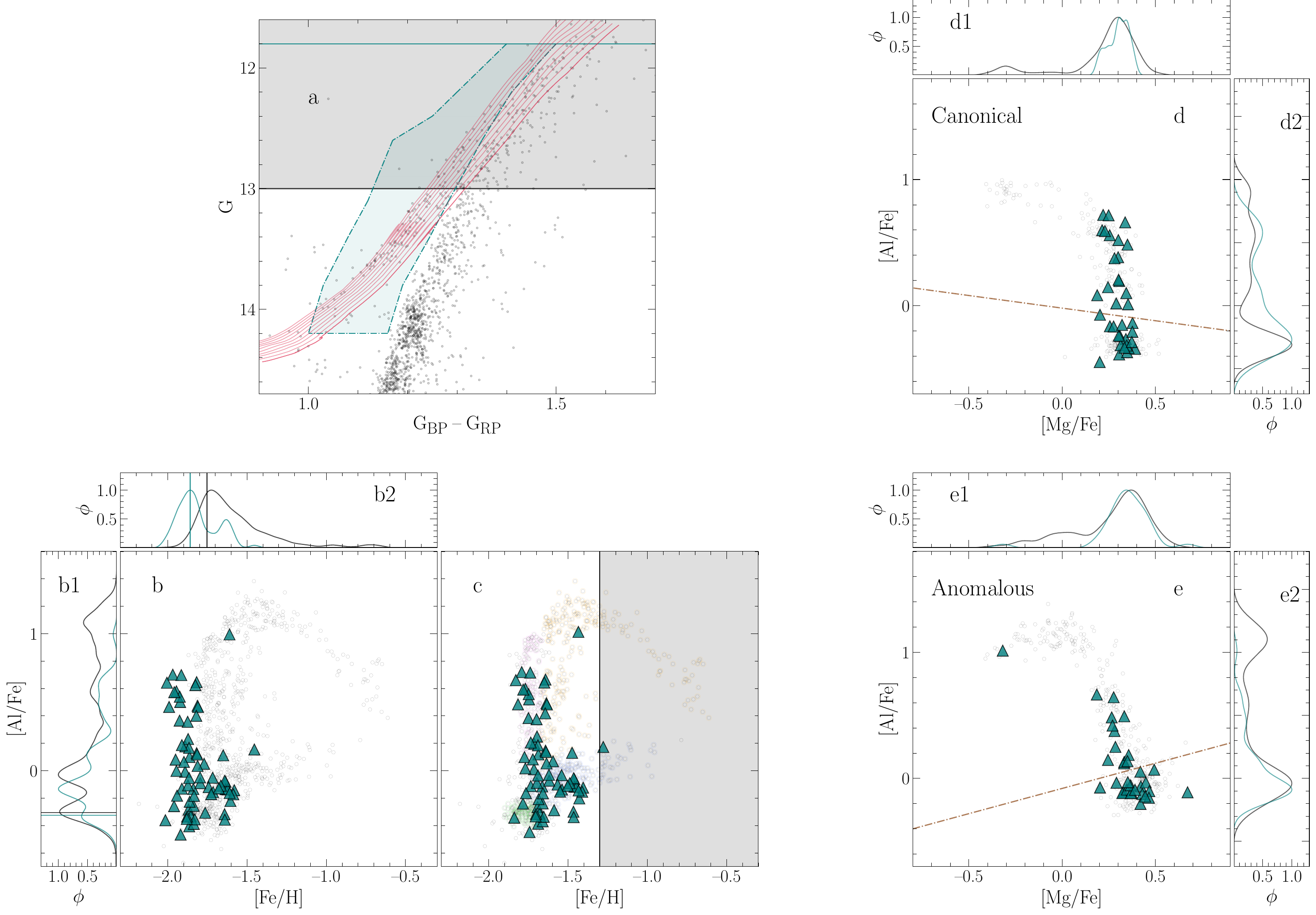}
\caption{
{\it{Panel a:}} $G$ vs. $G_{\rm BP}$--$G_{\rm RP}$ CMD of $\omega$Centauri. Black dots indicate RGB stars, while the selected AGB stars lie within the teal shaded area. Red lines indicate AGB isochrones from [Fe/H] --1.75 to --1.30 dex. The black line separates the magnitude range considered in our analysis from the one that we excluded (gray shaded area).
{\it{Panel b:}} [Al/Fe] vs. [Fe/H] of RGB (black dots) and AGB (teal triangles) below $G =$ 13 mag.
{\it{Panels b1 and b2:}} [Al/Fe] and [Fe/H] kernel density distributions, respectively, of RGB and AGB stars. Lines indicate the median 1P abundances of both group of stars.
{\it{Panel c:}} [Al/Fe] vs. [Fe/H] corrected for offsets. The black vertical line marks the [Fe/H]$=$--1.30 dex level, above which (gray shaded area) no RGB-AGB distiction is reliable in the Gaia CMD. Green, violet, blue, and orange dots indicate 1P, 2P, AI, and AII RGB stars as classified by \citet[][]{dondoglio2026}.
{\it{Panels d and e:}} [Al/Fe] vs. [Mg/Fe] of canonical and anomalous stars, respectively. Brown dot-dashed lines separates 1P from 2P and AI from AII.
{\it{Panels c1 and c2:}} [Mg/Fe] and [Al/Fe] kernel density distribution of canonical RGB and AGB stars.
{\it{Panels d1 and d2:}} same as panels c1 and c2 but for the anomalous stars.}
\label{fig:wcen_abn}
\end{figure*}

Owing to the large iron variations, the identification of AGB stars in the $G$ vs. $G_{\rm BP}$--$G_{\rm RP}$ CMD is particularly challenging, as the more metal-rich AGB stars overlap with the RGB sequence. To construct a self-consistent and unambiguous AGB sample, we begin by selecting by eye a bona fide CMD area occupied by AGB as done in Figure~\ref{fig:sel_agb}, represented by a teal shaded area in panel a of Figure~\ref{fig:wcen_abn}.
We overlay the AGB of Basti $\alpha$-enhanced isochrones \citep[][]{hidalgo2018} with $E(B-V) = $ 0.13 mag and age of 12.08 Gyr, which represents the mean age derived by \citet[][]{clontz2024}\footnote{The age distribution of $\omega$ Centauri remains a matter of debate, with different studies supporting total age spreads ranging from a few hundred Myr to several Gyr \citep[e.g.,][]{tailo2016, clontz2024}. In Appendix~\ref{sec:ap2}, we demonstrate that even in the presence of large age variations, our AGB selection remains unaffected.}, and [Fe/H] ranging from --1.75 (corresponding to the most metal-poor stars in the APOGEE dataset) up to --1.30 dex, indicated by red lines. Below $G = 13$ mag (black horizontal line), the AGB isochrones in this [Fe/H] range remain clearly separated from the RGB region of the CMD, while at brighter magnitudes they overlap significantly with the RGB, with only the most metal-poor ones falling in the teal region, thus making a reliable classification impossible. For this reason, we restrict our analysis to stars fainter than $G = 13$ mag. This approach ensures completeness in the selection of anomalous AGB stars up to [Fe/H] --1.30 dex, while more iron-rich anomalous stars cannot be safely isolated from the RGB, and are therefore excluded from our forthcoming analysis.

Panel~b of Figure~\ref{fig:wcen_abn} shows the [Al/Fe] vs. [Fe/H] diagram for RGB and AGB stars (fainter than $G = 13$ mag), which was used by \citet[][]{dondoglio2026} to separate canonical and anomalous populations, as well as their respective subpopulations. As in Section~\ref{sec:3}, we exploit the chemically homogeneous 1P stars to correct for possible RGB-to-AGB abundance biases. Dondoglio and collaborators demonstrated that, along the RGB, 1P stars populate a distinct, Al-poor clump centered at [Al/Fe] $\sim-0.3$ dex, which forms a distinct peak in the [Al/Fe] kernel density distribution in panel b1. Analogously, we identify 1P AGB stars as those forming the most Al-poor peak in the AGB distribution. The median [Al/Fe] of RGB and AGB are indicated with black and teal horizontal lines, respectively, from which we derive an RGB-to-AGB systematic offset of approximately 0.02 dex, in line with the values inferred for the other GCs in our sample.
A clear offset is also present in [Fe/H], as shown by its kernel density distributions in panel~b2, where AGB stars are more metal-poor than RGB stars. Systematically lower iron abundances among the AGB have been documented multiple times in the literature when, as in the APOGEE dataset, such quantities are derived from Fe I lines \citep[e.g.,][]{ivans2001, lapenna2014}.
To quantify this effect, we compute the median [Fe/H] values of RGB and AGB 1P stars, indicated by the vertical lines in panel~b2. The difference between these medians reveals AGB stars being $\sim$0.18 dex more Fe-poor than their RGB counterparts. We explore [Fe/H] offsets among our whole GCs sample in Appendix~\ref{sec:ap3}.

Panel~c shows the corrected [Al/Fe] vs. [Fe/H], where the gray shaded region marks the excluded [Fe/H] range beyond which RGB and AGB stars cannot be reliably distinguished across the full magnitude interval considered, as established in panel~a. The two evolutionary phases exhibit markedly different behaviors in the Al-rich regime, where AGB stars are essentially absent, especially in the Fe-rich regime (excluding the gray area), thus suggesting a stronger under-representation of Al-rich among anomalous stars. To tag the different populations in the RGB, we rely on the identification performed by \citet[][see their Figure 3]{dondoglio2026}, who disentangled the canonical 1P and 2P, and the anomalous AI and AII RGB stars (colored in green, violet, blue, and orange, respectively). The same population tagging is performed for AGB stars, based on which area of the [Al/Fe] vs. [Fe/H] diagram they occupy.

Panels~d and~e show the [Al/Fe] vs. [Mg/Fe] diagrams for canonical and anomalous stars, respectively. Among canonical stars, the Mg–Al anticorrelation within 1P and 2P stars (below and above the brown line, respectively) indicates that AGB stars follow the RGB trend up to [Mg/Fe]$\sim$0.2 dex and [Al/Fe]$\sim$0.8 dex, beyond which 2P stars are absent, showing a strong AGB-manqué footprint among the extreme 2P of $\omega$Centauri, as also supported by the kernel distributions portrayed in panels~d1 and ~d2. As well documented in the literature, also the anomalous population displays a similar Mg-Al anticorrelation \citep[e.g.,][]{mason2026}, with AII (above the brown line) ranging wider [Mg/Fe] and [Al/Fe] interval than AI (below the brown line), which forms the most Mg-rich and Al-poor clump of stars.
As observed in NGC\,6656, the AI dominates the AGB anomalous population, with a much larger incidence than in the RGB, as highlighted by the kernel distributions in panels~e1 and ~e2. Only a very small amount of AII stars populate the AGB, with just one of them having [Mg/Fe]$<$0.1 dex and [Al/Fe]$>$0.8 dex.

\section{Fraction of multiple populations: RGB vs. AGB}
\label{sec:4}

In this section, we measure the multiple population fractions in our stellar sample, with particular focus into comparing the RGB and AGB evolutionary phases.

We derive the fractions of 1P and 2P in the RGB ($F_{\rm 1P, 2P}^{\rm RGB}$) by counting the number of stars associated to each population as defined in Section~\ref{sec:3}. We then repeat the procedure in the AGB sequence to obtain the fraction of 1P and 2P among AGB stars ($F_{\rm 1P, 2P}^{\rm AGB}$).
The resulting fractions are reported in Table~\ref{tab:classify}. This represents the widest sample of GCs for which 1P and 2P fractions among AGB stars have been measured to date.
We estimate the uncertainty by using the Wilson score interval, adopting the 68.27\% confidence level. This choice is motivated by the robustness of the Wilson interval for small-sample statistics  -- especially relevant for the AGB stars -- that follow a binomial distribution
\citep[see the discussion by][]{brown2001}\footnote{DOI: \href{https://doi.org/10.1214/ss/1009213286}{10.1214/ss/1009213286}.}.

\begin{figure}
\includegraphics[width=7.0cm, clip, trim={ -1.0cm 0cm 0cm 0cm}]{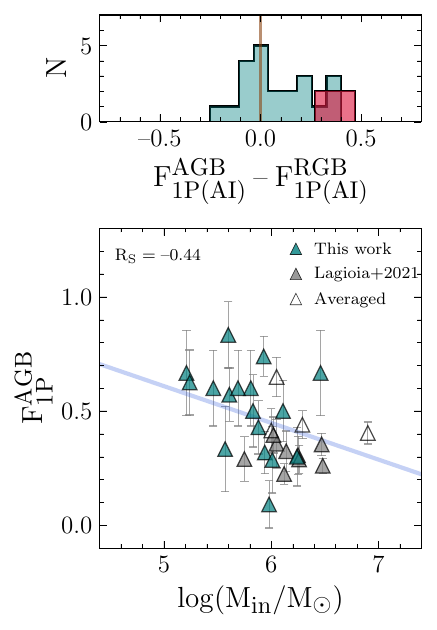}
\caption{
{\it{Top:}} differences between the 1P-fractions on the AGB and the RGB for Type I clusters and for canonical stars of Type II clusters (teal), and difference of AI-fractions for the anomalous stars of Type II clusters (red).
The brown vertical line indicates the zero level.
{\it{Bottom:}} $F_{\rm 1P}^{\rm AGB}$ vs log$(M_{\rm in}/M_{\rm \odot})$ from this work, from \citet[][]{lagioia2021}, and averaging the two datasets (teal, gray, and empty triangles, respectively).}
\label{fig:histogram}
\end{figure}

In the upper panel of Figure~\ref{fig:histogram}, we present the histogram of the difference between $F_{\rm 1P}^{\rm AGB}$ and $F_{\rm 1P}^{\rm RGB}$. The distribution peaks around zero, thus showing that in most of our GCs no significant differences in AGB and RGB 1P fraction is detected. Nevertheless, we point out that the distribution is skewed towards positive values, indicating that, when deviations from zero are present, a larger $F_{\rm 1P}^{\rm AGB}$ is the more typical case. Notable examples are NGC\,0288, NGC\,1904, NGC\,5053, NGC\,6171, NGC\,6205, NGC\,6218, and NGC\,6809, where $F_{\rm 1P}^{\rm AGB} > F_{\rm 1P}^{\rm RGB}$ at 1-$\sigma$ level.
By repeating our analysis on anomalous subpopulations, we quantify the AI and AII fraction of the anomalous populations in NGC\,6656 and $\omega$Centauri, confirming the visual impression reported in Sections~\ref{sec:3.2} and~\ref{sec:3.3} that the fraction of AII stars dramatically drops among the AGB, halving compared to their fraction in the RGB phase (as reported in Table~\ref{tab:classify}). Indeed, their histogram distribution (in red) in Figure~\ref{fig:histogram} reveals that the AI fraction among AGB stars is larger than in the RGB by almost 0.5. Intriguingly, this happens in both clusters, thus pointing to a shared behavior among anomalous stars in different Type II GCs.

The lower panel illustrates $F_{\rm 1P}^{\rm AGB}$ compared to the host GC initial mass \citep[from][]{baumgardt2018}. Here, we combined the fraction derived in this work for 22 GCs (teal triangles), with the fraction from \citet[][gray triangles]{lagioia2021} derived for 13 GCs. For the four clusters in common, we averaged the two estimates by weighting according to the number of AGB stars employed in each measurement (black empty triangles). A total of 31 GCs is therefore represented in the lower panel.
$F_{\rm 1P}^{\rm AGB}$ exhibits a mild anticorrelation with the initial GC mass (with Spearman's correlation coefficient valuing --0.44), also suggested by the decreasing trend highlighted by the best-fit line represented in blue. This is a typical behavior of multiple populations in GCs, and has been observed in other evolutionary phases, such as the RGB and HB \citep[][]{milone2017, dondoglio2021, lagioia2025}. For the first time, we show that this behavior is shared also among AGB stars.

In a simplified picture, the AGB-manqué phenomenon should act against the decreasing trend of $F_{\rm 1P}^{\rm AGB}$ with cluster mass: more massive GCs host the most He-rich 2P stars, which are expected to skip the AGB phase. Our results indicate a more complex situation: NGC\,2808 is a massive GC with large He enhancement, with only a small fraction of the most He-rich stars lacking in the AGB, while the low-mass, metal-poor, and poorly He-enriched NGC\,5053 exhibits a relatively stronger AGB-manqué signature. NGC\,6171 further illustrates that the phenomenon occurs in relatively metal-rich, low-mass systems. In addition, NGC\,6121 and NGC\,2808 show that the effect may be confined to the most extreme 2P stars, leading to only minor differences between RGB and AGB fractions. Overall, the AGB-manqué phenomenon depends on multiple parameters (He content, metallicity, HB morphology, and RGB mass loss) and likely additional second-order effects, and therefore does not systematically reverse the global trend with cluster mass.

\begin{figure}
\includegraphics[width=7.5cm, clip, trim={0cm 0.8cm 0cm 0.2cm}]{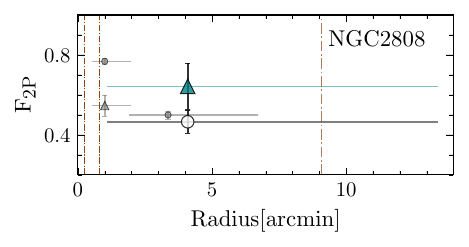}
\includegraphics[width=7.5cm, clip, trim={0cm 0.8cm 0cm 0.2cm}]{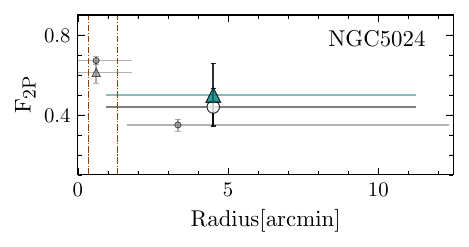}
\includegraphics[width=7.5cm, clip, trim={0cm 0.8cm 0cm 0.2cm}]{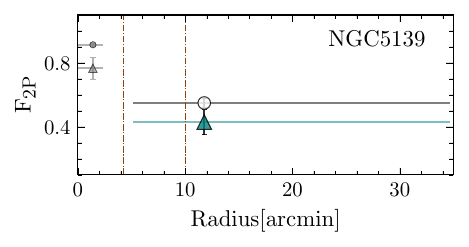}
\includegraphics[width=7.5cm, clip, trim={0cm 0cm 0cm 0.2cm}]{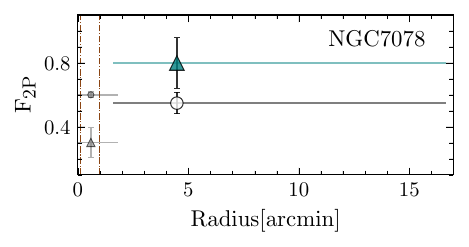}
\caption{$F_{\rm 2P}$ vs. radius in NGC\,2808, NGC\,5024, $\omega$Centauri, and NGC\,7078. RGB and AGB measurements from this work are represented with empty black dots and teal triangles. Filled grey dots and triangles indicate RGB and AGB fractions from the literature. The brown vertical lines indicate the core, half-mass, and tidal radius from \citet[][]{harris1996}. An horizontal line is associated to each measurement, highlighting their radial coverage.}
\label{fig:radial_2P}
\end{figure}

We now investigate the radial distribution of the 2P fraction among AGB stars. This analysis is possible for four GCs in our sample  -- NGC\,2808, NGC\,5024, $\omega$\,Centauri, and NGC\,7078 -- which overlap with the clusters studied by \citet[][]{lagioia2021}, who derived analogous quantities in the central regions using HST photometry. In Figure~\ref{fig:radial_2P}, we combine our measurements of the RGB and AGB 2P-fractions (shown as empty black circles and teal triangles, respectively) with literature estimates from photometric studies (gray symbols). These include RGB and AGB 2P-fractions measured in the cluster cores \citep[][]{milone2017, lagioia2021}, as well as RGB 2P-fractions in the outermost regions of NGC\,2808 \citep[from][]{jang2022} and NGC\,5024 \citep[from][]{leitinger2023}. For these two latter GCs, the outermost measurements cover the closest radial range to our estimates, showing a 1-$\sigma$ agreement with our $F_{\rm 2P}^{\rm RGB}$ values.
In $\omega$\,Centauri, we find a much smaller 2P fraction among RGB stars than the core estimate, which is expected knowing the large $F_{\rm 2P}^{\rm RGB}$ radial decrease \citep[see][]{dondoglio2026}.
In NGC\,7078, our $F_{\rm 2P}^{\rm RGB}$ is consistent with the core measurement even though it covers areas of the cluster further away from the center, thus suggesting the lack of a significant radial gradient among its population, as also found by \citet[][]{leitinger2023}.

Turning to the AGB population, different behaviors emerge. In NGC\,5024, the AGB 2P-fraction is consistent with the RGB values across all radial ranges. In contrast, $\omega$\,Centauri shows a systematically lower $F_{\rm 2P}^{\rm AGB}$ at all radii, in agreement with the AGB-manqué signature discussed in Section~\ref{sec:3.3}. A more complex pattern is observed in NGC\,2808 and NGC\,7078: while the AGB 2P-fraction is significantly lower than the RGB value in the central regions, it becomes larger than the RGB estimate at larger distances (approximately five times the half-mass radius in both clusters).
This behavior is unexpected and, to our knowledge, not predicted by current multiple population formation scenarios. Additional data will be required to increase the statistical robustness of our sample and determine whether this trend is a spurious effect or if reflects an intrinsic property of these clusters.
For completeness, the radial distributions of the remaining clusters in our sample, along with comparisons to literature RGB measurements, are presented in Appendix~\ref{sec:ap3}.

We conclude this section discussing the fraction of anomalous AGB stars ($F_{\rm AN}^{\rm AGB}$) in NGC\,6656 and $\omega$Centauri. \citet[][]{lagioia2021}, through HST multiband photometry, measured the fraction of anomalous stars in eight Type II GCs, finding that in six of them (including the two in common with our study) the anomalous fraction among the AGB is significantly smaller than in the RGB. Our findings are the following:

\begin{itemize}
    \item In NGC\,6656, along the RGB the anomalous fraction values $F_{\rm AN}^{\rm RGB} =$0.36$\pm$0.05, consistent with the estimate in its central areas of $\sim$40\% from \citet[][]{milone2017} and of $\sim$37\% from \citet[][]{lee2020}. On the other hand, we find $F_{\rm AN}^{\rm AGB} =$0.39$\pm$0.10, which is consistent with the RGB values but significantly different from the central estimate among AGB stars from \citep[][]{lagioia2021}, of about the 10\%.

    \item In $\omega$Centauri, we find $F_{\rm AN}^{\rm RGB} =$0.59$\pm$0.02, again consistent with the core estimate by \citet[][]{milone2017}. In the AGB, we find a slightly lower value (0.51$\pm$0.06), suggesting a smaller anomalous incidence. Still, our fraction is larger than the core AGB value of $\sim$30\% provided by \citet[][]{lagioia2021}. Our cut of selecting only anomalous stars richer than [Fe/H] $=$--1.30 dex may explain this discrepancy. Indeed, we notice in Figure~\ref{fig:wcen_abn} that the majority of anomalous AGB stars is clustered in the low-metallicity end compared to RGB stars. This suggests that at higher [Fe/H] the incidence of AGB is smaller, possibly reducing $F_{\rm AN}^{\rm AGB}$ to approach the fraction derived from Lagioia and collaborators, which indeed find no AGB stars among the metal richest populations in this cluster (see their Figure 16). To further corroborate this idea, we notice that in Figure~\ref{fig:wcen_abn} -- in the fainter end of the teal region where AGB and RGB exhibit a net separation -- AGB stars are almost non-existing at colors redder than the most metal-rich isochrone. The lack of stars in this area qualitatively suggests that in the metal-rich end of $\omega$Centauri, the contribution of AGB stars could be negligible.
\end{itemize}

\section{AGB-manqué: when and how}
\label{sec:5}

Figures~\ref{fig:mg_al} reveals that in some of our GCs the most chemically extreme 2P stars are missing along the AGB, suggesting the occurrence of the AGB-manqué phenomenon. In this Section, we investigate this feature in detail, providing a quantitative criterion to define which GCs exhibit AGB-manqué signatures.

\begin{figure}
\includegraphics[height=4.6cm, clip, trim={ 0cm 0cm 8cm 0cm}]{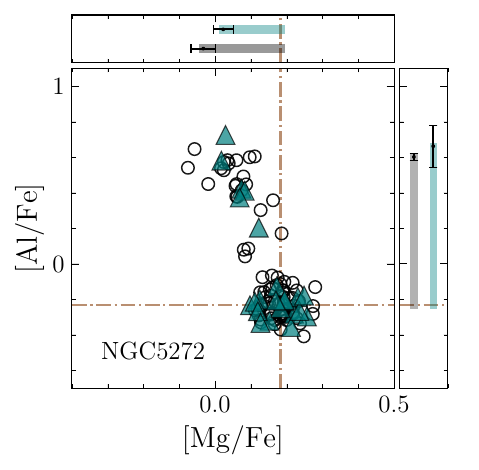}
\includegraphics[height=4.6cm, clip, trim={ 0.9cm 0cm 8cm 0cm}]{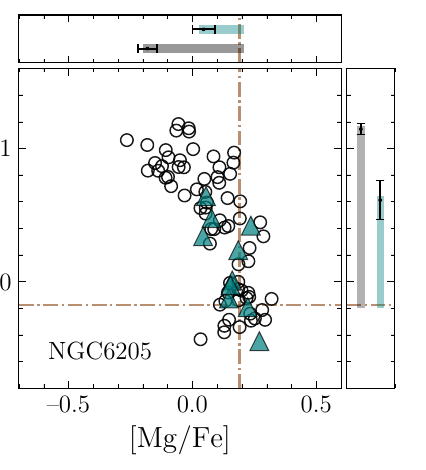}
\caption{
[Al/Fe] vs. [Mg/Fe] of NGC\,5272 (left) and NGC\,6205 (right). Brown lines indicate the median 1P abundances. Both plots are associated with a top and right panel, representing the RGB and AGB extent with gray and teal shaded bands, respectively. Black points show the 98th percentile of each distribution (see text for details).}
\label{fig:manqué_ex}
\end{figure}

\begin{figure*}
\includegraphics[width=16.5cm, clip, trim={ 0cm 0cm 0cm 18.1cm}]{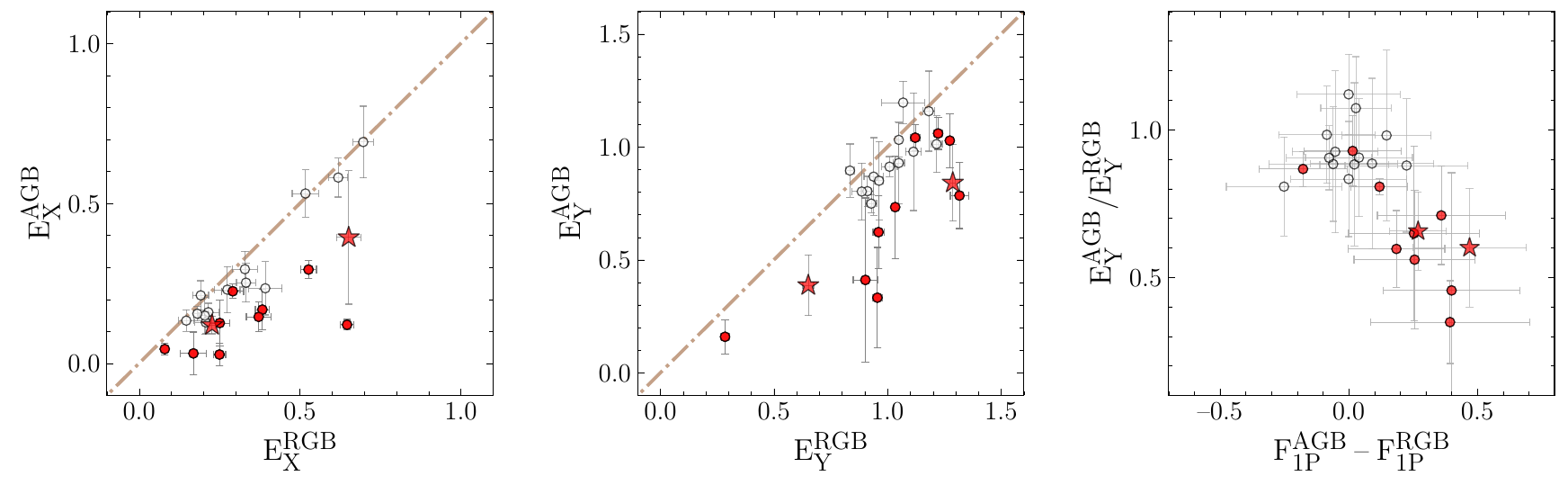}
\caption{AGB vs RGB extent of X$=$[Mg/Fe], [C/Fe]$_{\rm STD}$ (left), and Y$=$[Al/Fe], [N/Fe]$_{\rm STD}$ (middle). The brown dot-dashed line indicates the one-on-one relation. The right panel portrays the ratio between $E_{\rm Y}^{\rm AGB}$ and $E_{\rm Y}^{\rm RGB}$ vs. the AGB-to-RGB 1P fraction difference. Red dots indicate the GCs with AGB-manqué signatures, while red starred symbols refer to the anomalous populations.}
\label{fig:extension}
\end{figure*}

To identify clusters with AGB-manqué imprints, we adopt the procedure illustrated in Figure~\ref{fig:manqué_ex}, which shows [Al/Fe] vs. [Mg/Fe] for NGC\,5272 (left) and NGC\,6205 (right), extracted from Figure~\ref{fig:mg_al}. The brown lines mark the median 1P abundances in each cluster.
For both the RGB and AGB, we quantify the [Mg/Fe] and [Al/Fe] extent as the 98th percentile of the distance between the 1P median values and the whole 1P+2P stellar distribution. To account for observational uncertainties, the contribution of abundances errors to the observed spread is subtracted in quadrature. The resulting extents, $E_{\rm [Mg/Fe]}$ and $E_{\rm [Al/Fe]}$, are shown in the upper and right panels associated with each [Al/Fe] vs. [Mg/Fe] diagram. Gray and teal shaded regions represent RGB and AGB stars, respectively, while black points mark the 98th-percentile values. The associated uncertainties are estimated via bootstrap resampling (1,000 realizations with replacement), including also the contribution introduced by the offset correction on AGB stars.
As a quantitative criterion to classify a GC as affected by this phenomenon, we require that the AGB extents are smaller than the RGB ones in both quantities, i.e. $E_{\rm [Mg/Fe]}^{\rm AGB} < E_{\rm [Mg/Fe]}^{\rm RGB}$ and $E_{\rm [Al/Fe]}^{\rm AGB} < E_{\rm [Al/Fe]}^{\rm RGB}$, with the difference significant at $1\sigma$. NGC\,5272 and NGC\,6205 exhibit markedly different behaviors. In NGC\,5272, AGB stars span abundance ranges comparable to the RGB. In contrast, NGC\,6205 shows a significantly reduced AGB extent in both [Mg/Fe] and [Al/Fe], indicating a lack of the most chemically extreme stars along the AGB. Indeed, according to our definition, NGC\,5272 is not affected by the phenomenon, whereas NGC\,6205 shows clear evidence for the presence of the AGB-manqué phenomenon.

We apply the same procedure to the full cluster sample (using [C/Fe]$_{\rm STD}$ and [N/Fe]$_{\rm STD}$ for the metal-rich GCs). Based on this criterion, nine GCs display the AGB-manqué features: NGC\,0288, NGC\,1904, NGC\,2808, NGC\,4590, NGC\,5053, $\omega$Centauri, NGC\,6121, NGC\,6171, and NGC\,6205. In Figure~\ref{fig:extension}, we compare the AGB and RGB extents measured along the abundance axis used to identify multiple populations in each cluster, namely X = [Mg/Fe] (or [C/Fe]$_{\rm STD}$ for the metal-rich GCs; left panel) and Y = [Al/Fe] (or [N/Fe]$_{\rm STD}$; central panel). Thus, each point represents the comparison between the RGB and AGB extents derived using the same abundance indicator, rather than a comparison between [Mg/Fe] and [C/Fe]$_{\rm STD}$ (or between [Al/Fe] and [N/Fe]$_{\rm STD}$). Clusters highlighted in red lie below the one-on-one relations (brown line), and are those classified as undergoing the AGB-manqué phenomenon. We report the measured extents in Table~\ref{tab:classify}.
Our finding that NGC\,0288, NGC\,2808, NGC\,6121, and NGC\,6205 host AGB manqué is in agreement with previous spectroscopic investigation \citep[][]{campbell2012, johnson2012, wang2017, marino2019} and photometric studies \citep[][]{castelli2006, lagioia2021}. We provide the first spectroscopic evidence of $\omega$Centauri lacking its most extreme 2P stars, a result supported by the photometric analysis of \citet[][]{lagioia2021}. To our knowledge, this is the first time that AGB-manqué stars signatures have been detected in NGC\,1904, NGC\,4590, NGC\,5053, and NGC\,6171. 
The lack of AGB-manqué stars footprint detected in 47Tucanae, NGC\,5024, NGC\,5272, NGC\,6254, NGC\,6341, NGC\,6397, NGC\,6809, and NGC\,6838 in our work is in agreement with previous findings in the literature \citep[e.g., ][]{smith1989, gruyters2017, wang2017, gerber2018, maclean2018, lagioia2021}, while in NGC\,3201, NGC\,6218, NGC\,6656, and NGC\,7078 we provide the first ever spectroscopic identification of multiple populations among AGB stars.
The only GC with a strong discrepancy with the literature is NGC\,7078, in which we do not observe a lack of AGB stars among the 2P, contrary to the photometric analysis from \citet[][]{lagioia2021}, as shown by the large $F_{\rm 1P}^{\rm AGB}$ difference. A possible origin may reside in the peculiar AGB radial behavior detected in Figure~\ref{fig:radial_2P}, since the two analyses are based on different radial intervals.

In the right panel of Figure~\ref{fig:extension}, we show the AGB-to-RGB ratio of the measured extent along Y against the 1P fraction difference from Figure~\ref{fig:histogram}.
As expected, most of the manqué GCs occupy the part the region with the smallest extent ratio and the largest 1P fraction difference, while the other GCs are clustered around ($F^{\rm AGB}_{\rm 1P}$--$F^{\rm RGB}_{\rm 1P}$, $E^{\rm AGB}_{\rm Y}/E^{\rm AGB}_{\rm Y}$)$\sim$(0, 1). The only exceptions are NGC\,2808, the canonical stars of $\omega$Centauri, and NGC\,6121, where only a small portion of 2P stars skip the AGB phase, and therefore the differences compared to the RGB multiple populations pattern are smaller than in other GCs.

\begin{figure*}
\includegraphics[width=8.0cm, clip, trim={ 0cm 9cm 18.5cm 0cm}]{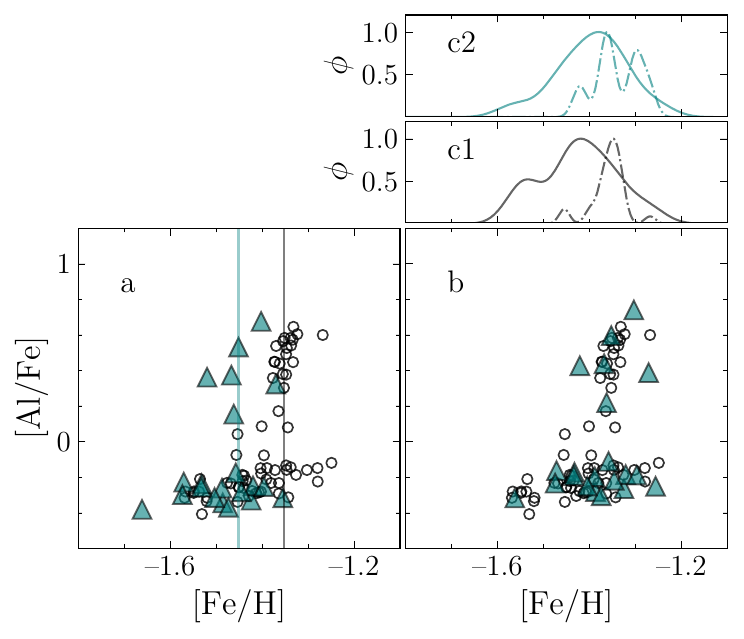}
\includegraphics[width=8.0cm, clip, trim={ 20cm 13cm 0cm 0cm}]{FIGURES/Sec_6/IRON_NGC5272.pdf}
\caption{
{\it{Panel a:}} [Al/Fe] vs. [Fe/H] of RGB (black dots) and AGB (teal triangles) of NGC\,5272. Black and teal vertical lines indicate their median [Fe/H] values, respectively.
{\it{Panel b:}} same as panel a, but corrected for RGB-to-AGB offsets.
{\it{Panels c1 and c2:}} kernel density distribution of 1P (continuous) and 2P (dot-dashed) in the RGB and AGB phase, respectively.
{\it{Panel d:}} radial distribution of $W_{\rm [Fe/H]}^{\rm 1P}$ combining our measurements (open dots) with the estimate by \citet[][black-filled dot]{legnardi2022}. Brown dashed line indicate the core and half-light radius, while the gray horizontal line indicate the radial coverage of each measurement.}
\label{fig:iron_1p}
\end{figure*}

We now extend this analysis to the anomalous populations in NGC\,6656 and $\omega$ Centauri, where the AGB-manqué occurrence is already suggested by the lack of the most Mg-poor and Al-rich AII AGB stars (Figures~\ref{fig:ngc6656} and~\ref{fig:wcen_abn}). Consistently, both clusters satisfy our criteria for exhibiting AGB-manqué imprints among AII, as shown in Figure~\ref{fig:extension}, where the abundance extents for anomalous stars are marked with red starred symbols.
Notably, in both systems the RGB-to-AGB discrepancy in the [Al/Fe] extent increases when moving from the canonical to the anomalous. This trend is consistent with the results presented in Section~\ref{sec:4}, where we found that the fraction of AII stars decreases more sharply along the AGB than that of 2P stars.
Overall, both clusters exhibit a recurring pattern: a pronounced paucity of AII stars along the AGB, particularly at the Mg-poor and Al-rich end. This provides evidence that anomalous populations are more severely affected by the AGB-manqué occurrence than the canonical populations within the same cluster.
Possible causes may reside in an increased He enhancement and/or mass loss -- two effects that are well known to increase the likelihood of a star to skip the AGB -- among AII stars compared to the 2P. Supporting this idea, \citet[][]{clontz2025} recently measured the He enhancement at different [Fe/H] in $\omega$Centauri, finding that in the metal-poor end (i.e., where canonical 1P and 2P stars lie), the maximum enhancement is lower than at higher metallicities (see their Figure 5). Studies devoted to measure He enrichment among anomalous stars in NGC\,6656 and other Type II GCs are a natural next step in unveiling this intriguing phenomenon.

\section{Discovery of an extended 1P on the AGB} \label{sec:6}

Recent studies have revealed that 1P stars exhibit a small but significant spread in photometric diagrams that cannot be explained by observational uncertainties. Comparisons between stellar models and photometry, together with high-resolution spectroscopy, indicate that intrinsic variations in [Fe/H] are the primary driver of this effect, commonly referred to as the extended 1P phenomenon \citep[e.g.,][]{marino2019, lardo2023, latour2025}.
NGC\,5272 is one of the GCs displaying the largest photometric extent of the 1P and, correspondingly, one of the widest [Fe/H] spreads observed spectroscopically \citep[][]{legnardi2022, dondoglio2025}. Moreover, we have [Fe/H] abundances for a relative large sample of 23 1P AGB stars, thus making it an ideal target to investigate whether iron variations are also present during this evolutionary phase.

In panel~a of Figure~\ref{fig:iron_1p}, we show the [Al/Fe] versus [Fe/H] for RGB and AGB stars. As shown in Figure~\ref{fig:wcen_abn}, [Fe/H] in the AGB is systematically lower than in the RGB. Given that we do not expect 1P stars to be homogeneous in their iron content, we exploit the 2P stars (which from panel~a clearly span a much smaller [Fe/H] range) to measure and correct for such effect. The median [Fe/H] difference between AGB and RGB (vertical lines) values around 0.1 dex.
Panel~b presents the corrected diagram, from which is evident that both RGB and AGB 1P span a wider [Fe/H] interval than the 2P stars. This can be appreciated from panel~c1, displaying the kernel density distributions of [Fe/H] for 1P (solid line) and 2P (dot-dashed line) RGB stars, and panel~c2, which illustrates the same but for AGB stars. 
1P stars span a clearly wider interval, ranging from $\sim$--1.55 to --1.25 dex.

To quantify the width of the 1P iron spread, $W_{\rm [Fe/H]}^{\rm 1P}$, we follow the approach outlined in \citet[][]{dondoglio2025}, which consists in measuring the difference between the 90th and the 10th percentile of the 1P [Fe/H] distribution. To this quantity, we then subtract in quadrature the contribution of the observational errors, measured through the difference between the 90th and the 10th percentile of a Gaussian distribution with standard deviation equal to the median [Fe/H] observational errors of our dataset. We find that $W_{\rm [Fe/H]}^{\rm 1P, RGB} = $0.23$\pm$0.03 and $W_{\rm [Fe/H]}^{\rm 1P, AGB} = $0.19$\pm$0.05 dex, thus proving that the 1P spread among RGB and AGB stars is the same within observational errors.

Finally, we explore the radial behavior of $W_{\rm [Fe/H]}^{\rm 1P}$. A similar investigation has been carried by \citet[][]{legnardi2024} by exploiting different photmetric diagrams at different radial distances for 47Tucanae, finding that the [Fe/H] spread of 1P stars slightly decreases moving outwards. In panel~d of Figure~\ref{fig:iron_1p}, we compare the iron-width estimate of NGC\,5272 from \citet[][]{legnardi2022} based on stars in the innermost two arcmin (gray dot), with the $W_{\rm [Fe/H]}^{\rm 1P}$ derived from our sample (by combining RGB and AGB stars for larger statistics) in two different, equal-number-of-stars, radial bins. We notice that our outermost estimate (within about 6 to 20 arcmin) is $\sim$0.05 dex smaller than the other estimates, hinting at a shared radial behavior between NGC\,5272 and 47Tucanae. However, the small $W_{\rm [Fe/H]}^{\rm 1P}$ difference lie within errorbars, thus lacking a robust statistical significance to asses if the trend is physical or error-driven.

Our analysis demonstrates that, in NGC\,5272 -- and potentially in other GCs -- the extended 1P phenomenon persists along the AGB. This represents the first detection of the extended 1P phenomenon in the AGB evolutionary phase.

\section{Conclusions}
\label{sec:7}

In this work, we demonstrate the power of combining Gaia photometry and APOGEE spectroscopy to probe the multiple-populations along the AGB, one of the least explored evolutionary phases in this context. By coupling light-element abundances with a rigorous AGB selection based on Gaia CMDs, we disentangle 1P, 2P, and (when present) anomalous stars in 22 Galactic GCs. Our main findings can be summarized as follows:

\begin{itemize}

\item The APOGEE [Al/Fe] vs. [Mg/Fe] and [N/Fe]$_{\rm STD}$ vs. [C/Fe]$_{\rm STD}$ diagrams are highly effective in separating 1P and 2P stars along the AGB. We provide population fractions for the largest GC sample analyzed to date in this evolutionary phase, showing that the 1P incidence decreases with increasing cluster mass, in agreement with results derived for other evolutionary stages \citep[e.g., ][]{milone2017}. Combined with the fractions derived by \citet[][]{lagioia2021}, we now have 1P and 2P ratios in the AGB for 31 GCs.

\item Our dataset enables a direct comparison between multiple-population along the RGB and AGB. We define a quantitative criterion to establish whether a GC displays the AGB-manqué phenomenon, finding that in nine GCs the most chemically extreme 2P stars -- those farthest from the 1P composition in light elements -- are underrepresented along the AGB relative to the RGB. This indicates that these stars failed to ascend the AGB, becoming AGB-manqué objects. In six of these nine clusters, the 1P fraction among AGB stars is significantly larger compared to the RGB, as expected if a portion of 2P stars is missing (in the other three, only a small subset of 2P skipped the AGB, thus not producing statistically significant fraction differences). The remaining clusters show AGB light-element extents and 1P fractions consistent within 1\,$\sigma$ with RGB values, indicating no evidence for AGB-manqués. Overall, our classification agrees with previous literature results, and provides the first spectroscopic characterization of AGB multiple populations in eight GCs (NGC\,1904, NGC\,4590, NGC 5053, NGC\,6171, NGC 3201, NGC\,6218, NGC\,6656, and NGC\,7078).

\item By combining our AGB multiple population fractions with estimates from \citet[][]{lagioia2021} in the central cluster areas, we derive for the first time the radial distribution of 2P AGB stars in four GCs. While in two of them (NGC\,5024 and $\omega$Centauri) $F_{\rm 2P}^{\rm AGB}$ follows a similar trend compared to the RGB, in NGC\,2808 and NGC\,7078 the AGB 2P stars exhibit an unexpected increase in their incidence moving outwards, an opposite behavior to what inferred from RGB stars. Dedicated spectroscopic studies with improved statistics will be crucial in testing and further exploring such findings.

\item We present the first spectroscopic mapping of anomalous AGB stars and their subpopulations in NGC\,6656 and $\omega$Centauri. In both clusters, the AGB-manqué signatures are particularly pronounced among the anomalous AII stars, especially at the Mg-poor and Al-rich extreme. The depletion of these stars along the AGB is even stronger than that observed among canonical 2P stars in the same clusters. We suggest that this behavior may reflect larger helium enrichment and/or enhanced RGB mass loss in these populations.

\item We report the first detection of iron inhomogeneities among 1P stars in the AGB phase for NGC\,5272. The measured width of the [Fe/H] distribution is consistent with that observed along the RGB, extending this phenomenon to the most evolved evolutionary phase studied so far, after previous detections in the MS, RGB, and HB \citep[e.g.,][]{marino2019, dondoglio2021, legnardi2024}. We also investigate the radial behavior of such iron spread. NGC\,5272 thus becomes only the second GC for which such an analysis is possible, after 47Tucanae \citep[][]{legnardi2024}. As in 47Tucanae, we find a mild decrease of the iron spread ($\sim$0.05 dex) toward larger radii. 

\end{itemize}

\begin{acknowledgements}
This work has been funded by the European Union – NextGenerationEU RRF M4C2 1.1 (PRIN 2022 2022MMEB9W: “Understanding the formation of globular clusters with their multiple stellar generations”, CUP C53D23001200006), from the European Union’s Horizon 2020 research and innovation programme under the Marie Skłodowska-Curie Grant Agreement No. 101034319, and from the European Union – NextGenerationEU (beneficiary: T. Ziliotto).
This work is supported by the FRG Grant and the Open Access Program from the American University of Sharjah. This paper represents the opinions of the authors and does not mean to represent the position or opinions of the American University of Sharjah. E. P. L. acknowledges support by Special Project for High-End Foreign Experts "Xingdian" Funding from Yunnan Province and National Key R\&D Program of China Grant (No. 2024YFA1611601).

\end{acknowledgements}

\bibliographystyle{aa}
\bibliography{aanda}

\begin{appendix}
\section{full sample information} \label{sec:ap1}

\begin{figure*}
\includegraphics[height=5.00cm, clip, trim={0cm 0cm 0cm 0cm}]{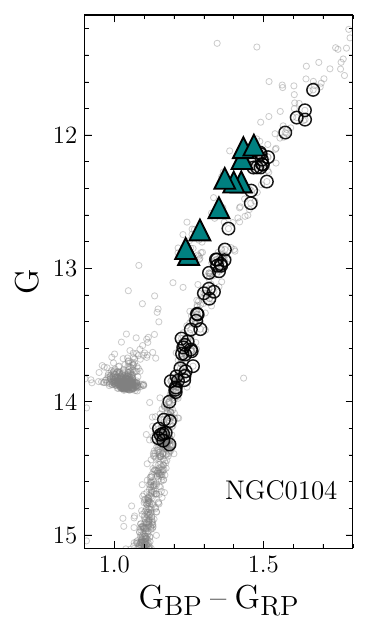}
\includegraphics[height=5.00cm, clip, trim={0.9cm 0cm 0cm 0cm}]{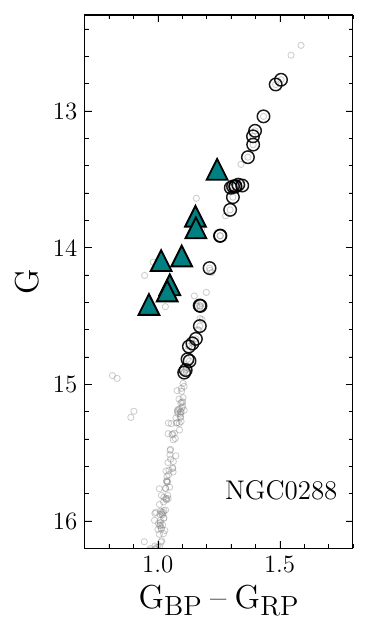}
\includegraphics[height=5.00cm, clip, trim={0.9cm 0cm 0cm 0cm}]{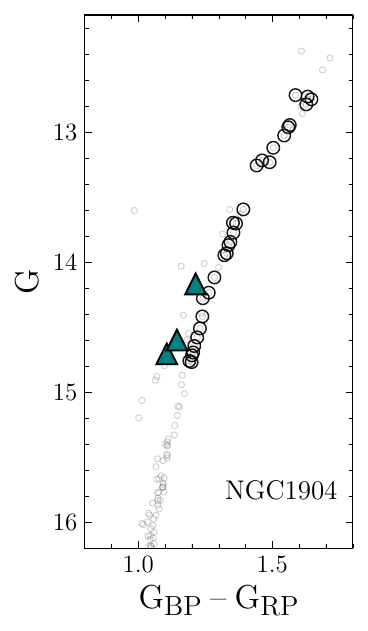}
\includegraphics[height=5.00cm, clip, trim={0.9cm 0cm 0cm 0cm}]{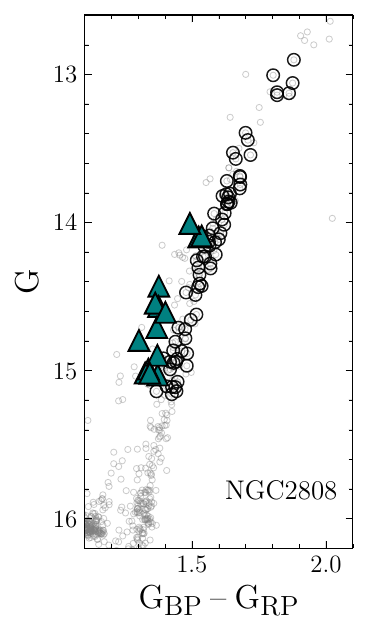}
\includegraphics[height=5.00cm, clip, trim={0.9cm 0cm 0cm 0cm}]{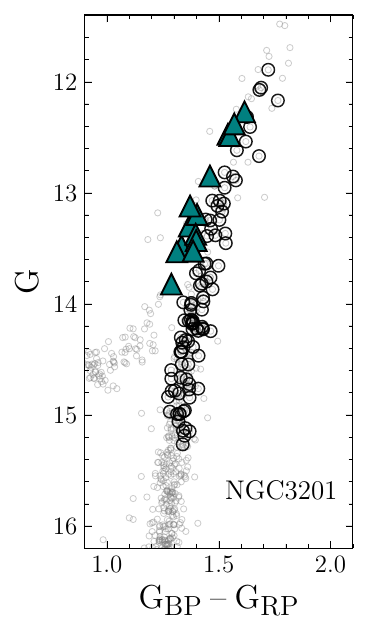}
\includegraphics[height=5.00cm, clip, trim={0.9cm 0cm 0cm 0cm}]{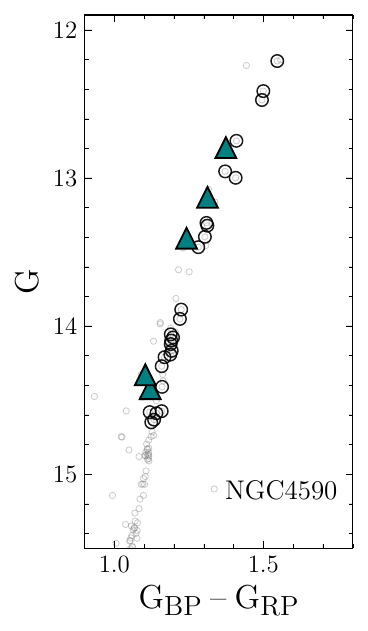}
\includegraphics[height=5.00cm, clip, trim={0.9cm 0cm 0cm 0cm}]{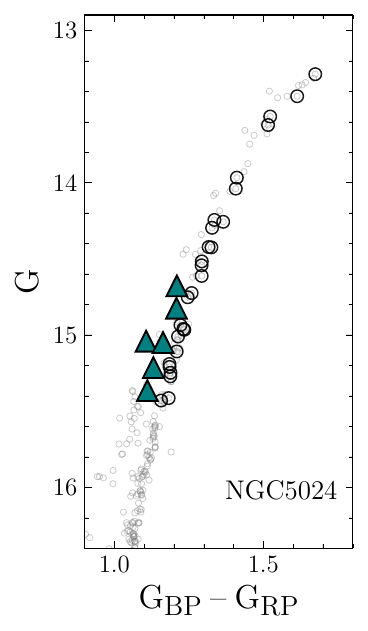}
\includegraphics[height=5.00cm, clip, trim={0cm 0cm 0cm 0cm}]{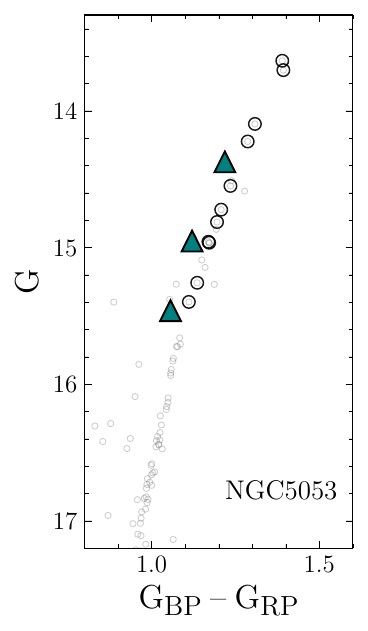}
\includegraphics[height=5.00cm, clip, trim={0.9cm 0cm 0cm 0cm}]{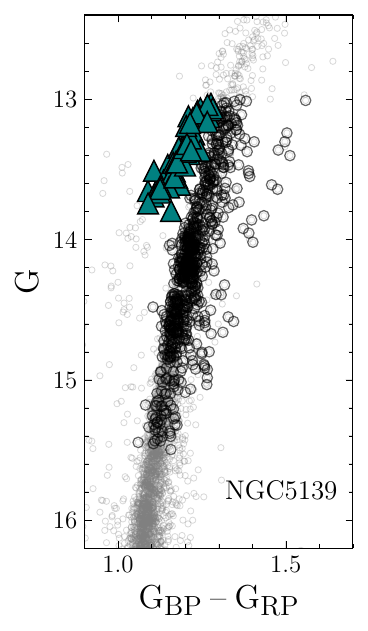}
\includegraphics[height=5.00cm, clip, trim={0.9cm 0cm 0cm 0cm}]{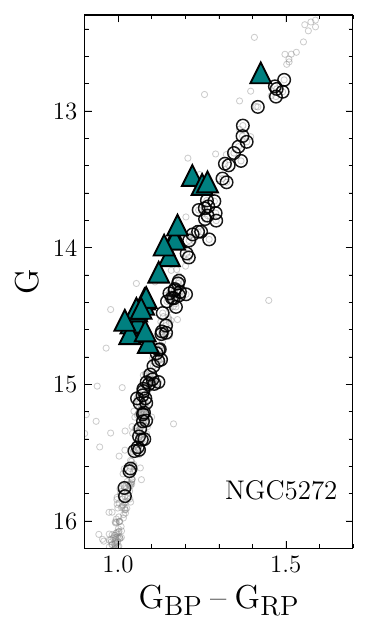}
\includegraphics[height=5.00cm, clip, trim={0.9cm 0cm 0cm 0cm}]{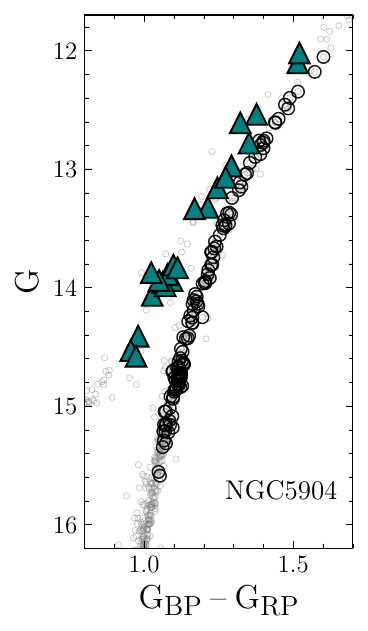}
\includegraphics[height=5.00cm, clip, trim={0.9cm 0cm 0cm 0cm}]{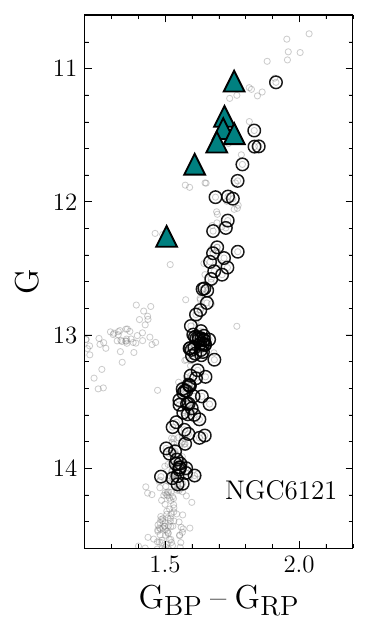}
\includegraphics[height=5.00cm, clip, trim={0.9cm 0cm 0cm 0cm}]{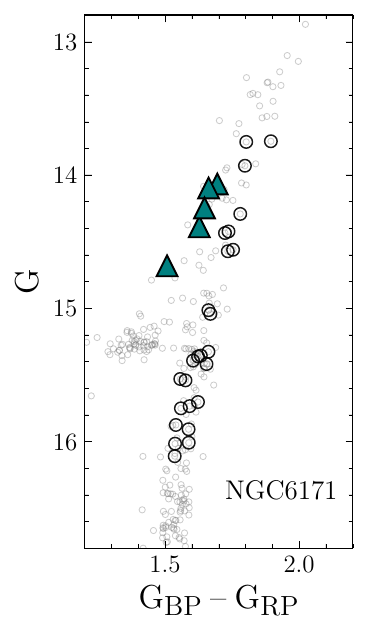}
\includegraphics[height=5.00cm, clip, trim={0.9cm 0cm 0cm 0cm}]{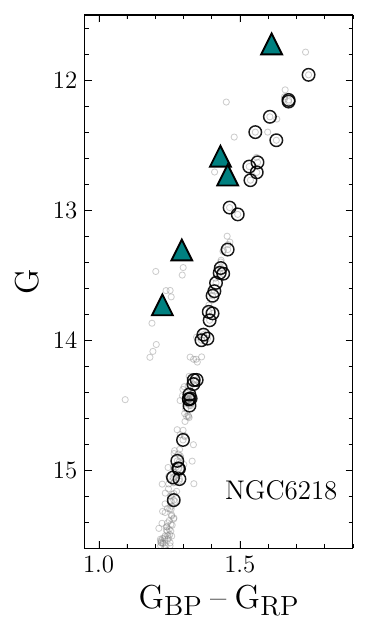}
\includegraphics[height=5.00cm, clip, trim={0.0cm 0cm 0cm 0cm}]{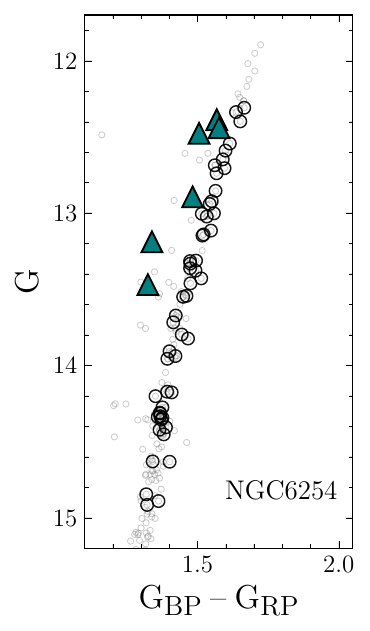}
\includegraphics[height=5.00cm, clip, trim={0.9cm 0cm 0cm 0cm}]{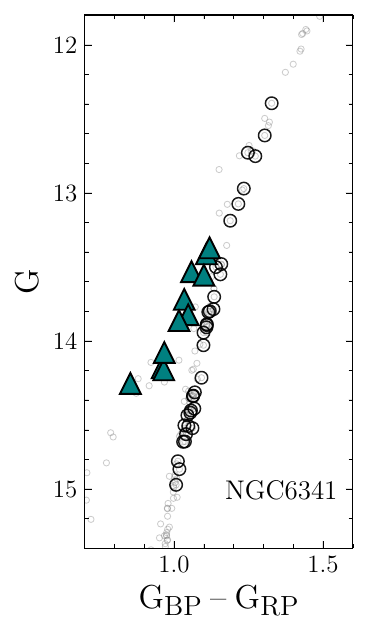}
\includegraphics[height=5.00cm, clip, trim={0.9cm 0cm 0cm 0cm}]{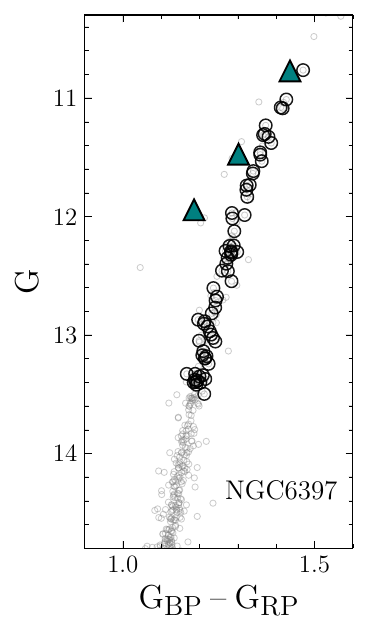}
\includegraphics[height=5.00cm, clip, trim={0.9cm 0cm 0cm 0cm}]{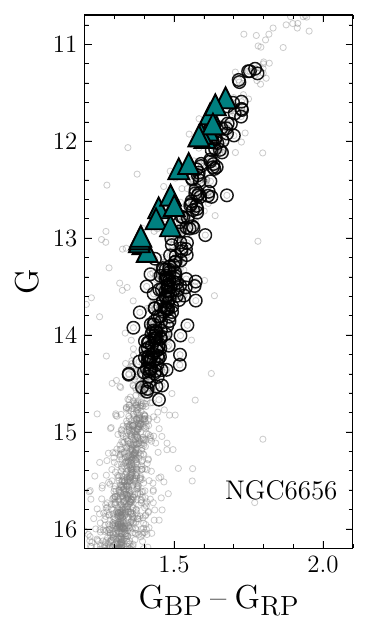}
\includegraphics[height=5.00cm, clip, trim={0.9cm 0cm 0cm 0cm}]{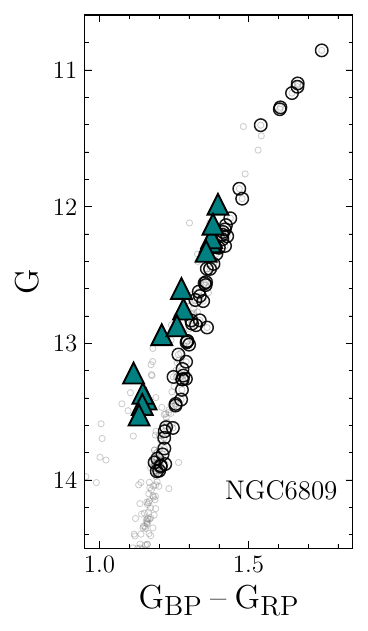}
\includegraphics[height=5.00cm, clip, trim={0.9cm 0cm 0cm 0cm}]{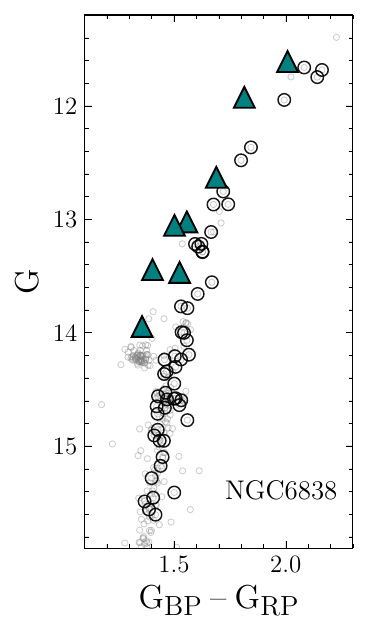}
\includegraphics[height=5.00cm, clip, trim={0.9cm 0cm 0cm 0cm}]{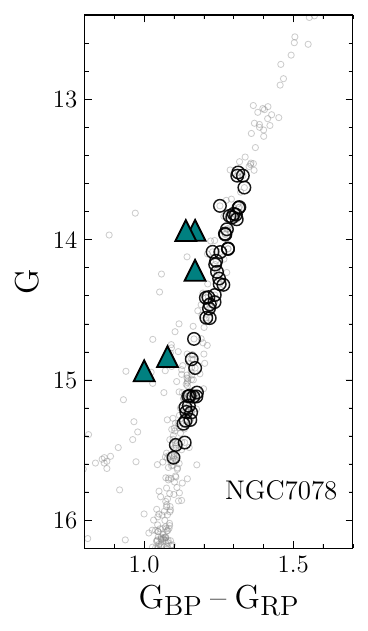}
\caption{$G$ vs. $G_{\rm BP}$-$G_{\rm RP}$ CMD of the GCs considered in this work. Black dots and teal triangles indicate RGB and AGB stars, respectively, with available APOGEE chemical abundances.}
\label{fig:all_cmd}
\end{figure*}

In this Appendix, we show in Figure~\ref{fig:all_cmd} the collection of the CMDs of all the GCs considered in our work. We exclude NGC\,6205 as it is already displayed in Figure~\ref{fig:sel_agb}. For each GC, we illustrate the differential-reddening corrected Gaia CMD of cluster member (gray points) and the RGB and AGB stars with available APOGEE abundances (black dots and teal triangles, respectively).

In Table~\ref{tab:classify}, we report the main observational features measured in this work: the number of RGB and AGB stars composing our sample, the 1P fractions, and the X and Y axis extents of the abundance ratios exploited in to separate multiple populations in Section~\ref{sec:3} on both evolutionary phases. For the two Type II GCs in our sample, we also report the same quantities relative to their anomalous populations.

\FloatBarrier

\FloatBarrier

\begin{table*}
    \centering
\caption{Summary of our sample of GCs, providing the number of RGB and AGB stars considered in this work, the fraction of 1P and 2P, AI and AII stars, and the extents along the X (X$=$[Mg/Fe], [C/Fe]$_{\rm STD}$) and Y (Y$=$[Al/Fe], [N/Fe]$_{\rm STD}$) axis of the diagram used in Section~\ref{sec:3}, in both evolutionary phases. Lines with the cluster name followed by '--A' include the anomalous subpopulations only, while the remainder are relative to the canonical 1P and 2P only.}
\begin{tabular}{l|c|c|c|c|c|c|c|c}
    \hline
    \hline
                  &        &        &                        &                        &                                           \\
     Cluster      & \# RGB & \# AGB & F$_{\rm 1P (AI)}^{\rm RGB}$ & F$_{\rm 1P (AI)}^{\rm AGB}$ & E$_{\rm X}^{\rm RGB}$ & E$_{\rm X}^{\rm AGB}$ & E$_{\rm Y}^{\rm RGB}$ & E$_{\rm Y}^{\rm AGB}$ \\
                  &        &        &                        &                        &                       &                       &                       &                       \\
    \hline												                                                                              
                  &        &        &                        &                        &                       &                       &                       &                       \\
    NGC\,0104     &  65    & 10     & 0.38 $\pm$ 0.06        & 0.30 $\pm$ 0.13        & 0.52 $\pm$ 0.04       & 0.53 $\pm$ 0.07       & 1.05 $\pm$ 0.01       & 1.03 $\pm$ 0.07       \\
    NGC\,0288     &  26    &  6     & 0.58 $\pm$ 0.09        & 0.83 $\pm$ 0.14        & 0.08 $\pm$ 0.01       & 0.04 $\pm$ 0.02       & 0.28 $\pm$ 0.02       & 0.16 $\pm$ 0.07       \\
    NGC\,1904     &  30    &  3     & 0.27 $\pm$ 0.08        & 0.67 $\pm$ 0.19        & 0.24 $\pm$ 0.05       & 0.12 $\pm$ 0.06       & 1.21 $\pm$ 0.10       & 0.68 $\pm$ 0.31       \\
    NGC\,2808     &  73    & 14     & 0.53 $\pm$ 0.06        & 0.36 $\pm$ 0.11        & 0.53 $\pm$ 0.03       & 0.29 $\pm$ 0.02       & 1.22 $\pm$ 0.02       & 1.06 $\pm$ 0.06       \\
    NGC\,3201     &  93    & 14     & 0.62 $\pm$ 0.05        & 0.57 $\pm$ 0.12        & 0.15 $\pm$ 0.02       & 0.13 $\pm$ 0.03       & 0.94 $\pm$ 0.04       & 0.87 $\pm$ 0.17       \\
    NGC\,4590     &  26    &  5     & 0.35 $\pm$ 0.09        & 0.60 $\pm$ 0.17        & 0.25 $\pm$ 0.03       & 0.13 $\pm$ 0.04       & 0.96 $\pm$ 0.02       & 0.62 $\pm$ 0.16       \\
    NGC\,5024     &  26    &  6     & 0.56 $\pm$ 0.09        & 0.50 $\pm$ 0.16        & 0.21 $\pm$ 0.03       & 0.13 $\pm$ 0.02       & 0.96 $\pm$ 0.02       & 0.85 $\pm$ 0.13       \\
    NGC\,5053     &  11    &  3     & 0.27 $\pm$ 0.12        & 0.67 $\pm$ 0.19        & 0.25 $\pm$ 0.02       & 0.03 $\pm$ 0.01       & 0.96 $\pm$ 0.02       & 0.33 $\pm$ 0.11       \\
    NGC\,5139     & 258    & 32     & 0.45 $\pm$ 0.03        & 0.57 $\pm$ 0.08        & 0.65 $\pm$ 0.02       & 0.12 $\pm$ 0.01       & 1.27 $\pm$ 0.01       & 1.03 $\pm$ 0.04       \\
    NGC\,5139$-$A & 366    & 33     & 0.46 $\pm$ 0.03        & 0.73 $\pm$ 0.08        & 0.65 $\pm$ 0.04       & 0.39 $\pm$ 0.21       & 1.28 $\pm$ 0.01       & 0.84 $\pm$ 0.17       \\
    NGC\,5272     &  83    & 23     & 0.71 $\pm$ 0.05        & 0.74 $\pm$ 0.09        & 0.22 $\pm$ 0.03       & 0.16 $\pm$ 0.02       & 0.83 $\pm$ 0.02       & 0.90 $\pm$ 0.11       \\
    NGC\,5904     & 125    & 22     & 0.30 $\pm$ 0.04        & 0.32 $\pm$ 0.09        & 0.19 $\pm$ 0.02       & 0.21 $\pm$ 0.03       & 0.91 $\pm$ 0.03       & 0.80 $\pm$ 0.07       \\
    NGC\,6121     & 100    &  7     & 0.27 $\pm$ 0.04        & 0.29 $\pm$ 0.15        & 0.29 $\pm$ 0.03       & 0.23 $\pm$ 0.02       & 1.12 $\pm$ 0.02       & 1.04 $\pm$ 0.04       \\
    NGC\,6171     &  25    &  5     & 0.24 $\pm$ 0.08        & 0.60 $\pm$ 0.17        & 0.38 $\pm$ 0.02       & 0.17 $\pm$ 0.03       & 1.03 $\pm$ 0.01       & 0.73 $\pm$ 0.22       \\
    NGC\,6205     &  70    & 10     & 0.31 $\pm$ 0.05        & 0.50 $\pm$ 0.13        & 0.37 $\pm$ 0.04       & 0.15 $\pm$ 0.02       & 1.32 $\pm$ 0.04       & 0.78 $\pm$ 0.09       \\
    NGC\,6218     &  39    &  6     & 0.38 $\pm$ 0.07        & 0.60 $\pm$ 0.17        & 0.40 $\pm$ 0.05       & 0.20 $\pm$ 0.04       & 1.12 $\pm$ 0.03       & 0.97 $\pm$ 0.25       \\
    NGC\,6254     &  52    &  6     & 0.50 $\pm$ 0.07        & 0.50 $\pm$ 0.16        & 0.33 $\pm$ 0.03       & 0.25 $\pm$ 0.06       & 1.22 $\pm$ 0.03       & 1.01 $\pm$ 0.13       \\
    NGC\,6341     &  36    & 11     & 0.17 $\pm$ 0.06        & 0.09 $\pm$ 0.10        & 0.62 $\pm$ 0.03       & 0.58 $\pm$ 0.06       & 1.01 $\pm$ 0.03       & 0.91 $\pm$ 0.02       \\
    NGC\,6397     & 66    &  3     & 0.24 $\pm$ 0.05        & 0.33 $\pm$ 0.19        & 0.26 $\pm$ 0.03       & 0.15 $\pm$ 0.06       & 1.08 $\pm$ 0.02       & 0.93 $\pm$ 0.18       \\
    NGC\,6656     &  50    & 11     & 0.36 $\pm$ 0.07        & 0.36 $\pm$ 0.13        & 0.18 $\pm$ 0.01       & 0.16 $\pm$ 0.01       & 1.07 $\pm$ 0.10       & 1.20 $\pm$ 0.09       \\
    NGC\,6656$-$A &  22    &  7     & 0.39 $\pm$ 0.09        & 0.86 $\pm$ 0.13        & 0.23 $\pm$ 0.01       & 0.12 $\pm$ 0.03       & 0.65 $\pm$ 0.02       & 0.39 $\pm$ 0.13       \\
    NGC\,6809     &  64    & 14     & 0.28 $\pm$ 0.06        & 0.43 $\pm$ 0.12        & 0.27 $\pm$ 0.04       & 0.23 $\pm$ 0.07       & 1.18 $\pm$ 0.03       & 1.16 $\pm$ 0.18       \\
    NGC\,6838     &  53    &  8     & 0.58 $\pm$ 0.07        & 0.63 $\pm$ 0.14        & 0.33 $\pm$ 0.04       & 0.29 $\pm$ 0.05       & 0.89 $\pm$ 0.04       & 0.80 $\pm$ 0.12       \\
    NGC\,7078     &  51    &  5     & 0.45 $\pm$ 0.07        & 0.20 $\pm$ 0.16        & 0.70 $\pm$ 0.03       & 0.69 $\pm$ 0.11       & 0.93 $\pm$ 0.03       & 0.75 $\pm$ 0.03       \\
                  &        &        &                        &                        &                       &                       &                       &                       \\
    \hline
    \hline
\end{tabular}
\label{tab:classify}
\end{table*}

\section{abundance offsets between RGB and AGB stars} \label{sec:ap3}

In this Appendix, we further investigate the systematic abundance differences between RGB and AGB stars introduced in Section~\ref{sec:3}.

\begin{figure*}
\includegraphics[height=4.9cm, clip, trim={ 0cm 0cm 0cm 0cm}]{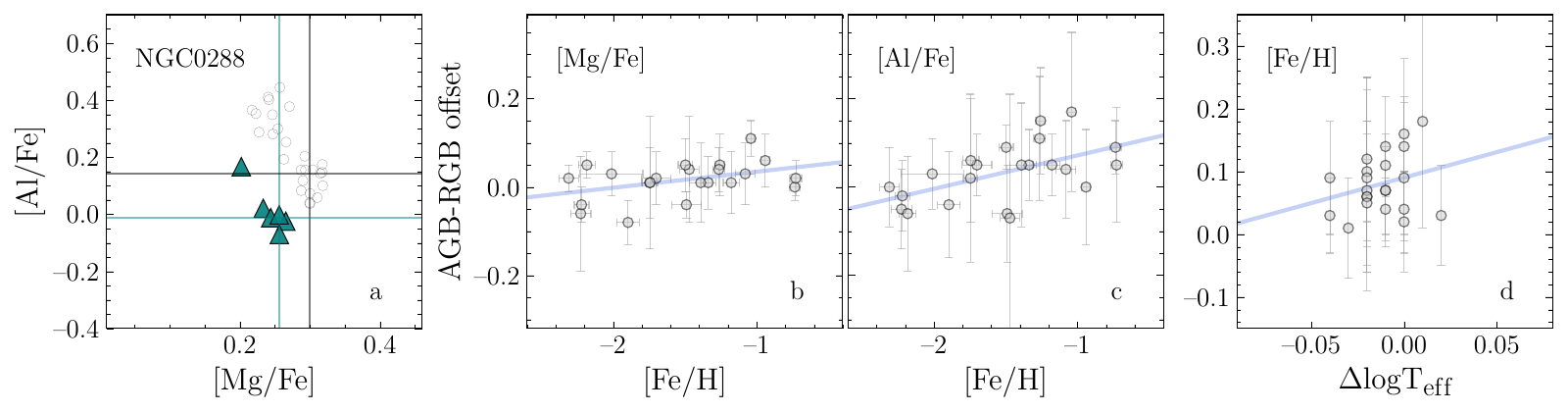}
\caption{{\it{Left:}} [Al/Fe] vs. [Mg/Fe] of RGB (black dots) and AGB (teal triangles) in NGC\,0288. Black and teal lines represent median 1P values of both groups of stars.
{\it{Middle:}} [Mg/Fe] and [Al/Fe] offsets against [Fe/H].
{\it{Right:}} [Fe/H] offsets vs. the logarithm of the average $T_{\rm eff}$ difference between AGB and RGB stars.
Best-fit lines are represented in blue.
}
\label{fig:systematics}
\end{figure*}

Panel~a of Figure~\ref{fig:systematics} illustrates the case of NGC\,0288, one of the clusters exhibiting the largest RGB-to-AGB abundance offsets in our sample. In this GC, AGB stars are systematically shifted toward lower [Mg/Fe] and [Al/Fe] values compared to the RGB population. As in Figure~\ref{fig:mg_al_6205}, black and teal lines indicate the median [Mg/Fe] and [Al/Fe] abundances of RGB and AGB 1P stars, respectively. We measure offsets of $\sim$0.05 dex in [Mg/Fe] and $\sim$0.15 dex in [Al/Fe].

Panels~b and~c present the RGB--AGB abundance offsets in [Mg/Fe] and [Al/Fe] for all the GCs in our sample, computed by considering only 1P stars. This choice minimizes potential biases introduced by the different multiple-population patterns among clusters. For the GCs shown in the last row of Figure~\ref{fig:mg_al}, where the 1P--2P separation cannot be reliably performed in the [Al/Fe]--[Mg/Fe] plane, we instead adopt the classification based on the [N/Fe]$_{\rm STD}$ versus [C/Fe]$_{\rm STD}$ diagram.
We find that the RGB--AGB offsets span ranges of $-0.08$ to $+0.07$ dex in [Mg/Fe] and $-0.07$ to $+0.15$ dex in [Al/Fe]. The offsets can therefore be either positive (AGB stars more abundance-poor than RGB stars) or negative (AGB stars more abundance-rich), and display a tendency to increase with cluster metallicity, particularly in the case of [Al/Fe], as also indicated by the positive slope of the best-fit relation shown in blue.

The median APOGEE uncertainties in [Mg/Fe] and [Al/Fe] for the analyzed stars are typically of the order of $\sim$0.03--0.05 dex. These values are compatible with the observed offsets only for a subset of the analyzed GCs, while several systems exhibit discrepancies significantly larger than the quoted uncertainties. Even under the conservative assumption that APOGEE abundance errors may be underestimated (given that they are often smaller than uncertainties reported in literature spectroscopic studies) an error-only explanation cannot account for the clear rigid shift observed in clusters such as NGC\,0288. We therefore conclude that these offsets are not purely driven by observational uncertainties. For this reason, and to ensure a homogeneous treatment across the full cluster sample, we apply the correction procedure described in Section~\ref{sec:3.1} even in cases where the measured offsets are comparable to, or smaller than, the formal abundance uncertainties (e.g., NGC\,6205 in Figure~\ref{fig:delta_n}).

The physical origin of these offsets remains uncertain, and a comprehensive investigation is beyond the scope of this work. Nevertheless, the observed increase of the offsets with [Fe/H] suggests that they may arise from RGB-to-AGB atmospheric differences that are not fully captured by the APOGEE Stellar Parameters and Chemical Abundances Pipeline (ASPCAP).

We perform the same analysis for [Fe/H]. In this case, the offsets are always positive, with AGB stars systematically more metal-poor than RGB stars by $\sim$0.02--0.19 dex. This effect has been widely reported in the literature when -- as in the APOGEE dataset -- iron abundances are derived from Fe~I lines \citep[e.g.,][]{ivans2001, lapenna2014}, and is commonly interpreted as a consequence of departures from local thermodynamic equilibrium (NLTE).
Unlike the Mg and Al offsets, the [Fe/H] discrepancy does not correlate with cluster metallicity. Instead, it shows a mild dependence on the difference in effective temperature between the RGB and AGB samples ($\Delta T_{\rm eff}$), illustrated in panel~d. $\Delta T_{\rm eff}$ has been derived by considering only 1P stars, and by considering only RGB stars that lie in the same $G$ range than AGB stars. Larger $\Delta T_{\rm eff}$ values correspond roughly to larger [Fe/H] offsets, supporting a scenario in which temperature-dependent NLTE effects participate in driving the observed discrepancies.

\FloatBarrier

\section{age effects on the AGB selection in $\omega$Centauri} \label{sec:ap2}

As anticipated in Section~\ref{sec:3.3}, the age distribution of the stellar populations in $\omega$Centauri constitutes a matter of debate. Indeed, while some works favor a scenario in which the cluster that we observe today took several Gyr to form \citep[e.g.,][]{villanova2014, clontz2024}, other support the idea that $\omega$Centauri is coeval within few hundreds of Myr \citep[][]{tailo2016, dondoglio2026}.
While shedding this mystery is not a goal of this study, we explore the influence that large age difference may have on our AGB stars selection procedure.

\begin{figure}
\includegraphics[width=8cm, clip, trim={ 9cm 35cm 25cm 0cm}]{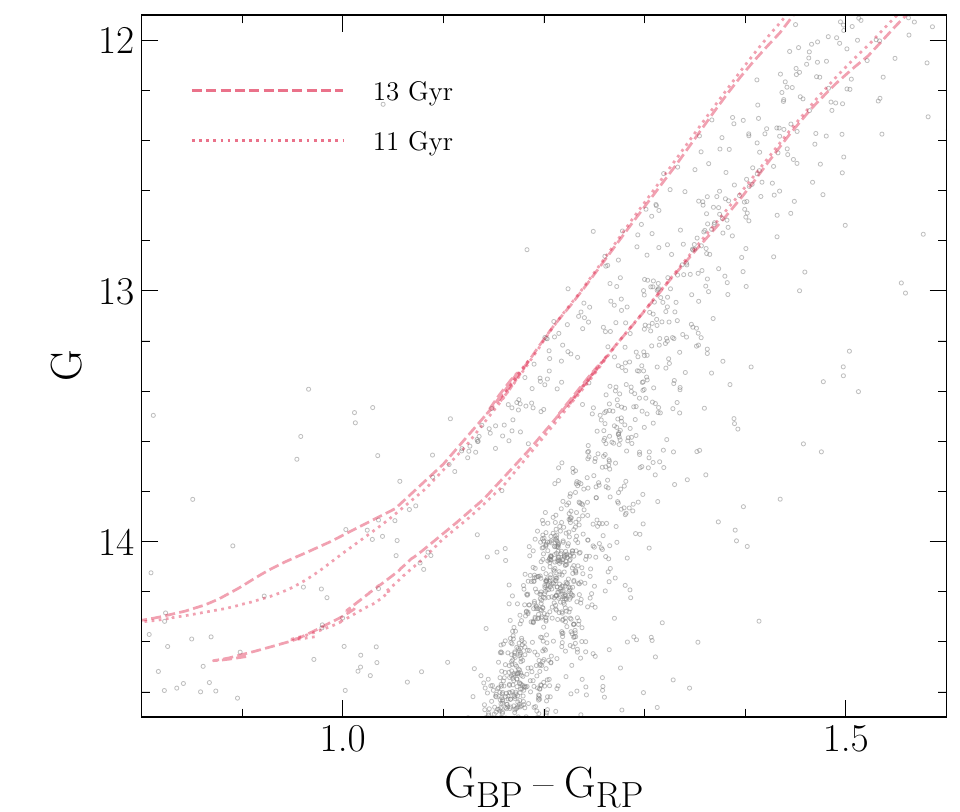}
\caption{$G$ vs. $G_{\rm BP}$--$G_{\rm RP}$ CMD of $\omega$Centauri. Red lines indicate AGB isochrones with [Fe/H] --1.75 to --1.30 dex, with ages of 11 (dotted lines) and 13 (dashed lines) Gyr.}
\label{fig:wcen_app}
\end{figure}

We show in Figure~\ref{fig:wcen_app} the $G$ vs. $G_{\rm BP}-G_{\rm RP}$ CMD as in panel a of Figure~\ref{fig:wcen_abn}, focusing on the most metal-poor and metal-rich isochrones ([Fe/H] $=-1.75$ and $-1.30$ dex) used to define the metallicity range over which our selection is reliable. For each metallicity, we compare 13 Gyr and 11 Gyr (which are roughly the mean age of the oldest and youngest populations measured by \citealt{clontz2024}, see their Figure~5) isochrones, shown as dashed and dotted lines, respectively.

In both metallicity regimes, isochrones of different ages but identical [Fe/H] almost perfectly overlap in the Gaia CMD. The resulting color differences remain below $\sim$0.01 mag along most of the AGB. This demonstrates that, at such old ages, the position of the AGB sequence in the adopted CMD is essentially insensitive to age variations across the full metallicity range spanned by our $\omega$ Centauri sample. Therefore, the use of a single representative age in Section~\ref{sec:3.3} is fully justified.

\FloatBarrier

\section{radial distribution} \label{sec:ap4}

\begin{figure*}
\includegraphics[height=3.7cm, clip, trim={0.00cm 0cm 0cm 0cm}]{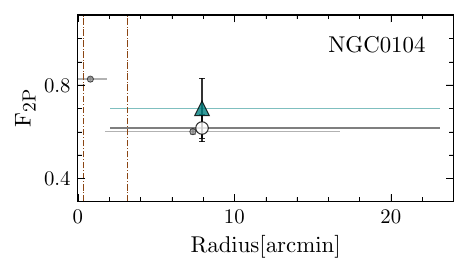}
\includegraphics[height=3.7cm, clip, trim={0.64cm 0cm 0cm 0cm}]{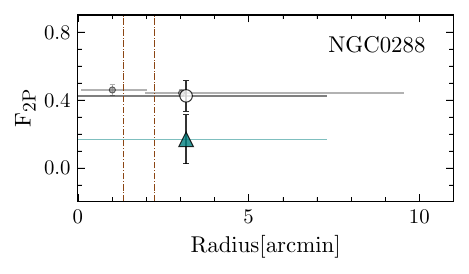}
\includegraphics[height=3.7cm, clip, trim={0.64cm 0cm 0cm 0cm}]{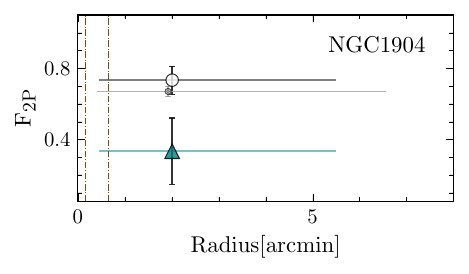}
\includegraphics[height=3.7cm, clip, trim={0.00cm 0cm 0cm 0cm}]{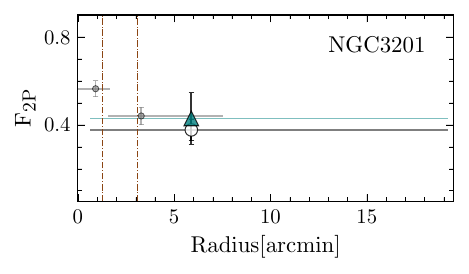}
\includegraphics[height=3.7cm, clip, trim={0.64cm 0cm 0cm 0cm}]{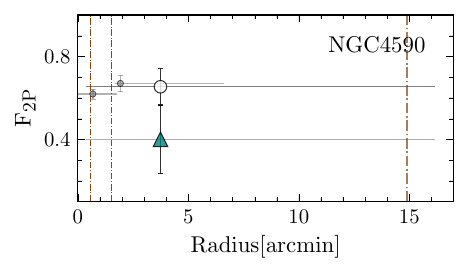}
\includegraphics[height=3.7cm, clip, trim={0.64cm 0cm 0cm 0cm}]{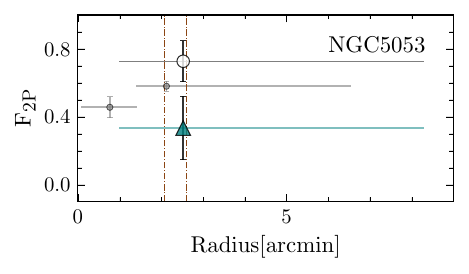}
\includegraphics[height=3.7cm, clip, trim={0.00cm 0cm 0cm 0cm}]{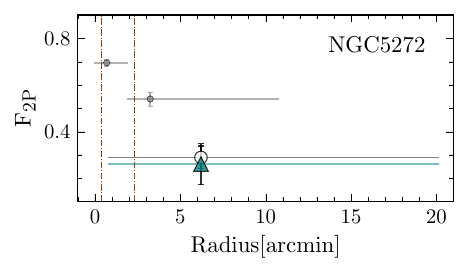}
\includegraphics[height=3.7cm, clip, trim={0.64cm 0cm 0cm 0cm}]{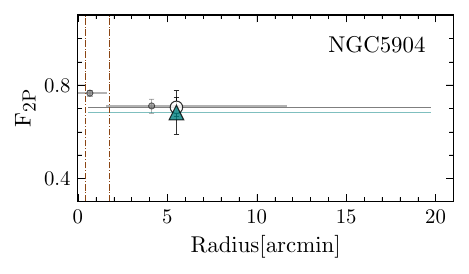}
\includegraphics[height=3.7cm, clip, trim={0.64cm 0cm 0cm 0cm}]{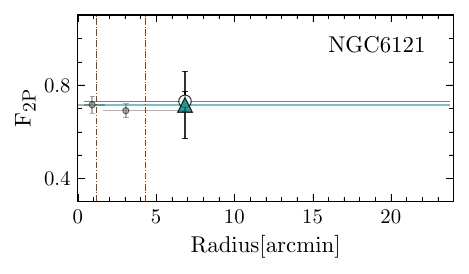}
\includegraphics[height=3.7cm, clip, trim={0.00cm 0cm 0cm 0cm}]{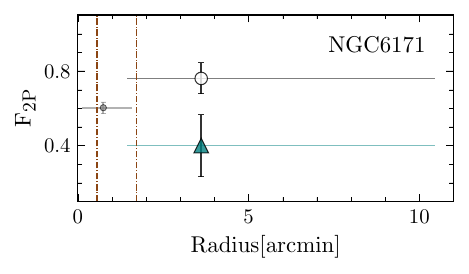}
\includegraphics[height=3.7cm, clip, trim={0.64cm 0cm 0cm 0cm}]{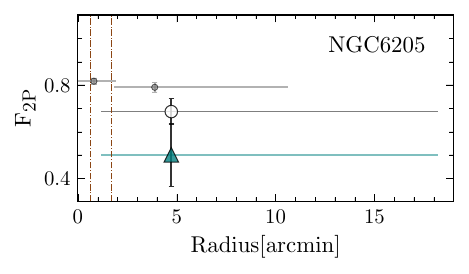}
\includegraphics[height=3.7cm, clip, trim={0.64cm 0cm 0cm 0cm}]{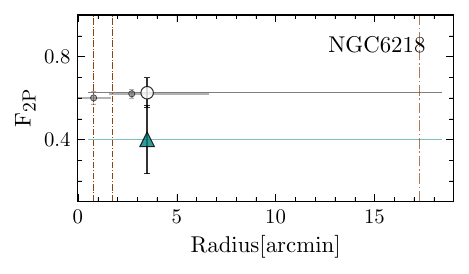}
\includegraphics[height=3.7cm, clip, trim={0.00cm 0cm 0cm 0cm}]{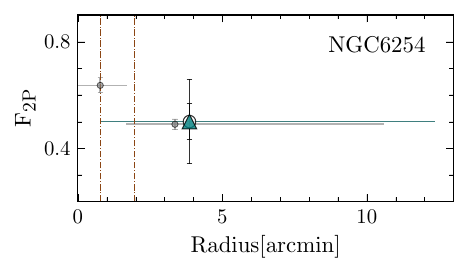}
\includegraphics[height=3.7cm, clip, trim={0.64cm 0cm 0cm 0cm}]{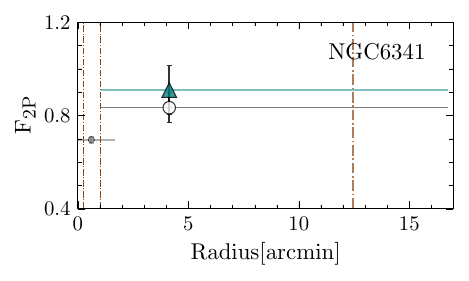}
\includegraphics[height=3.7cm, clip, trim={0.64cm 0cm 0cm 0cm}]{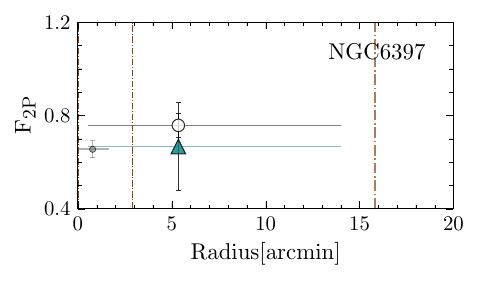}
\includegraphics[height=3.7cm, clip, trim={0.00cm 0cm 0cm 0cm}]{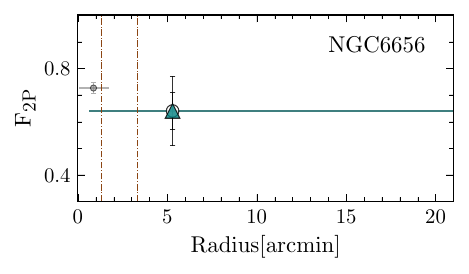}
\includegraphics[height=3.7cm, clip, trim={0.64cm 0cm 0cm 0cm}]{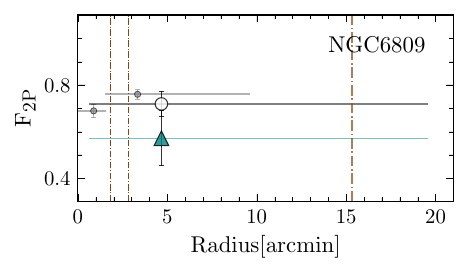}
\includegraphics[height=3.7cm, clip, trim={0.64cm 0cm 0cm 0cm}]{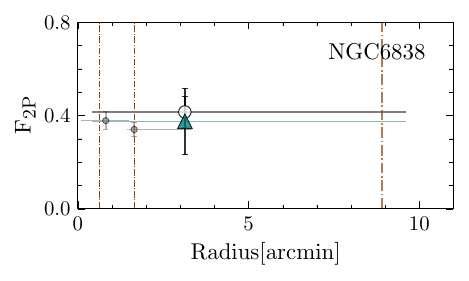}
\caption{Same as Figure~\ref{fig:radial_2P} but for the GCs not analysed in Section~\ref{sec:4}.}
\label{fig:radial}
\end{figure*}

We present in Figure~\ref{fig:radial} the radial distribution of the 2P fraction for all GCs in our sample (excluding clusters with core estimates from \citealt{lagioia2021}, illustrated in Figure~\ref{fig:radial_2P}) by comparing our estimates with photometric determinations from the literature, probing the innermost $\sim$2 arcmin \citep[from][]{milone2017} and the outermost cluster regions \citep[from][]{jang2025}.

Our RGB-based fractions confirm previously reported radial gradients in 47Tucanae, NGC\,3201, and NGC\,5272, where the 2P population becomes progressively less dominant at larger radii \citep[e.g.][]{lee2021, jang2025, metha2025}, consistent with a more centrally concentrated 2P component. This result is particularly relevant for NGC\,3201, whose radial distribution remains debated. While some studies \citep[][]{leitinger2023, cadelano2024} reported a more centrally concentrated 1P population, our findings support instead the results from \citet[][]{jang2025, metha2025}, favoring a centrally concentrated 2P.

When comparing our RGB 2P-fractions with literature measurements in similar radial ranges, we find overall consistency. In particular, 47Tucanae, NGC\,0288, NGC\,1904, NGC\,3201, NGC\,4590, NGC\,5053, NGC\,5904, NGC\,6121, NGC\,6205, NGC\,6218, NGC\,6254, NGC\,6809, and NGC\,6838 agree within 1-$\sigma$ with the estimates by \citet[][]{jang2025}. The only significant discrepancy is observed in NGC\,5272, where our 2P fraction is lower. This difference likely arises because our sample probes stars at larger median radii than those considered by Jang et al., where a reduced 2P fraction is expected given the strong radial decline reported for this cluster \citep[see also][]{lee2021}.
We also note that our RGB 2P-fractions for NGC\,5904 and NGC\,6341 are consistent with the values reported by \citet[][]{lee2017, lee2023}, whose analyses include stars outside the cluster cores.
In NGC\,6171, NGC\,6341, and NGC\,6397, we report a small but significant increase in the 2P fraction towards outer radii. However, the significance is slightly above 1-$\sigma$, therefore care must be taken before implying that this is a genuine 2P increase trend.

The AGB 2P-fractions derived in this work are consistent with the RGB-based values for 47Tucanae, NGC\,3201, NGC\,5272, NGC\,5904, NGC\,6121, NGC\,6254, NGC\,6341, NGC\,6397, NGC\,6656, NGC\,6809, and NGC\,6838. This agreement supports our previous conclusion that these clusters are not significantly affected by the AGB-manqué phenomenon and therefore exhibit similar multiple-population patterns along the RGB and AGB.
The only partial exception is NGC\,6121, which we identify as displaying AGB-manqué features despite showing comparable RGB and AGB 2P fractions. In this case, the phenomenon affects only a small portion of 2P stars (see Figure~\ref{fig:mg_al}), such that their absence does not produce a statistically significant difference in the overall 2P fraction, similarly to what is observed in NGC\,2808 (see Figure~\ref{fig:radial_2P}).

Conversely, the largest discrepancies between AGB and RGB 2P-fractions (differences of $\sim$0.2–0.4) are found in clusters where we detect clear evidence of AGB-manqués imprints, including NGC\,0288, NGC\,1904, NGC\,4590, NGC\,5053, NGC\,6171, and NGC\,6205.
An interesting case is NGC\,6218: although it does not formally satisfy our AGB-manqué criterion from Section~\ref{sec:5}, its AGB 2P-fraction is significantly smaller than the RGB value. This discrepancy is likely driven by small-number statistics, as only five AGB stars are available, leading to large uncertainties in $E_{\rm [C/Fe]_{\rm STD}}^{\rm AGB}$ and $E_{\rm [N/Fe]_{\rm STD}}^{\rm AGB}$. Indeed, while the [C/Fe]$_{\rm STD}$ extent is significantly reduced along the AGB, the same is not true for [N/Fe]$_{\rm STD}$. This mixed behavior suggests that NGC\,6218 may also be affected by the AGB-manqué phenomenon, although larger samples are required to confirm this possibility.

\end{appendix}
\end{document}